\documentclass[aps,pra,onecolumn,11pt]{revtex4}
\newcommand {\be}{\begin{equation}}
\newcommand {\ee}{\end{equation}}
\newcommand {\bey}{\begin{eqnarray}}
\newcommand {\eey}{\end{eqnarray}}
\newcommand {\true}{\emph{true}}
\newcommand {\false}{\emph{false}}
\newcommand {\bs}{\symbol{92}}
\usepackage{epsfig}
\usepackage{amsfonts}
\usepackage{amssymb}
\usepackage{amsmath}
\usepackage{mathbbol}
\usepackage{enumerate}
\usepackage{amsthm}
\usepackage{pgfplots}
\usepackage{tikz}

\usepackage{mathtools}

\usepackage{doi}
\usepackage{hyperref}

\newtheorem{definition}{Definition}
\newtheorem{theorem}{Theorem}
\theoremstyle{definition}
\newtheorem{problem}{Problem}
\newtheorem{lemma}{Lemma}
\newtheorem{corollary}{Corollary}
\newtheorem{algorithm}{Algorithm}
\newtheorem{property}{Property}
\newtheorem{claim}{Claim}
\newtheorem{procedure}{Procedure}

\begin{document}
\title{An Algebraic-Geometry Approach to Prime Factorization}
\author{Alberto Montina, Stefan Wolf}
\affiliation{Facolt\`a di Informatica, 
Universit\`a della Svizzera italiana, 6900 Lugano, Switzerland}
\date{\today}

\begin{abstract}
New algorithms for prime factorization that outperform the existing ones
or take advantage of particular properties of the prime factors can have 
a practical impact on present implementations of cryptographic algorithms 
that rely on the complexity of factorization. Currently used keys are chosen 
on the basis of the present algorithmic  knowledge and, thus, can potentially be 
subject to future breaches. For this reason, it is worth to investigate new approaches
which have the potentiality of giving a computational advantage. The problem has 
also relevance in quantum computation, as an efficient quantum algorithm for prime 
factorization already exists. Thus, better classical asymptotic complexity can provide 
a better understanding of the advantages offered by quantum computers. 
In this paper, we reduce the factorization problem to the search of points
of parametrizable varieties, in particular curves, over finite fields. The 
varieties are required to have an arbitrarily large number of intersection 
points with some hypersurface over the base field. 
For a subexponential or polynomial factoring
complexity, the number of parameters have to scale sublinearly in the space
dimension $n$ and the complexity of computing a point given the parameters has 
to be subexponential or polynomial, respectively.
We outline a procedure for building these varieties, which is
illustrated with two constructions. In one case, we show that there are varieties
whose points can 
be evaluated efficiently given a number of parameters not greater than $n/2$. In
the other case, the bound is dropped to $n/3$. Incidentally, the first construction 
resembles a kind of retro-causal model. Retro-causality is considered one possible 
explanation of quantum weirdness.
\end{abstract}

\maketitle

\section{Introduction}
\label{intro}
Prime factorization is a problem in the complexity class NP of problems that can be solved in 
polynomial time by a nondeterministic machine. Indeed, the prime factors  can be verified 
efficiently by multiplication.
At present, it is not known if the problem has polynomial computational complexity 
and, thus, is in the complexity class P. Nonetheless, the most common cryptographic algorithms 
rely on the assumption of hardness of factorization. Website certificates 
and bitcoin wallets are examples of resources depending on that
assumption. Since no strict lower bound on the computational complexity is actually
known, many critical services are potentially subject to 
future security breaches. Consequently, cryptographic keys have gradually increased
their length to adapt to new findings. For example, the general number-field sieve~\cite{nfsieve} 
can break keys that would have been considered secure against previous factoring methods.

Prime factorization is also important for its relation with quantum computing, since
an efficient quantum algorithm for factorization is known. This algorithm is considered
a main argument supporting the supremacy of quantum over classical computing. Thus,
the search for faster classical algorithms is relevant for better understanding the
actual gap between classical and quantum realm. 

Many of the known factoring methods use the ring of integers modulo $c$ as a common feature, where 
$c$ is the number to be factorized. Examples are Pollard's $\rho$ and $p-1$ 
algorithms~\cite{pollard1,pollard2}, Williams' $p+1$ algorithm~\cite{williams},
their generalization with cyclotomic polynomials~\cite{bach}, Lenstra elliptic curve 
factorization~\cite{lenstra}, and quadratic sieve~\cite{qsieve}.
These methods end up to generate a number, say $m$, having a 
factor of $c$. Once $m$ is obtained, the common factor can be efficiently 
computed by the Euclidean algorithm. Some of these methods use only operations defined in the ring. Others,
such as the elliptic-curve method, perform also division operations by pretending that $c$ is prime.
If this operation fails at some point, the divisor is taken as outcome $m$.
In other words, the purpose of these methods is to compute a zero over the field of integers modulo 
some prime factor of $c$ by possibly starting from some random initial state. Thus, the general scheme 
is summarized by a map $X\mapsto m$ from a random state $X$ in some set $\Omega$ to an integer $m$ 
modulo $c$. Different states may be tried until $m$ is equal to zero modulo some prime of $c$. 
The complexity of the algorithm depends on the computational complexity of generating $X$ in $\Omega$,
the computational complexity of evaluating the map, and the average number of trials  required to find a zero.

In this paper, we employ this general scheme by focusing on a class $\Theta$
of maps defined as multivariate rational 
functions over prime fields $\mathbb{Z}_p$ of order $p$ and, more generally, over a finite 
field $\text{GF}(q)$ of order $q=p^k$ and degree $k$. The set $\Omega$ of inputs is taken equal to
the domain of definition of the maps.  More precisely, the maps are first defined over some algebraic 
number field $\mathbb{Q}(\alpha)$ of degree $k_0$, where $\alpha$ is an algebraic number, that is, 
solution of some irreducible polynomial $P_I$ of degree $k_0$. Then, the maps
are reinterpreted over a finite field. 
Using the general scheme of the other methods, the class $\Theta$
takes to a factoring algorithm with polynomial complexity if 
\begin{enumerate}
\item[(a)] the number of distinct zeros, say $N_P$, of the maps in $\Theta$ is arbitrarily large over $\mathbb{Q}(\alpha)$;
\item[(b)] a large fraction of the zeros remain distinct when reinterpreted over a finite field
whose order is greater than about $N_P^{1/M}$;
\item[(c)] \label{cond_c}
the product of the number of parameters by the field degree is upper-bounded by a
sublinear power function of $\log N_P$;
\item[(d)] the computational complexity of evaluating the map given any input is upper-bounded 
by a polynomial function in $\log N_P$.
\end{enumerate}
A subexponential factoring complexity is achieved with weaker scaling conditions on 
the map complexity, as discussed later in Sec.~\ref{sec_complex}.
Later, this approach to factorization will be reduced to the search of rational points of a 
variety having an arbitrarily large number of rational intersection points with a hypersurface.

The scheme employing rational functions resembles some existing methods, such as Pollard's $p-1$ 
algorithm. The main 
difference is that these algorithms generally rely on algebraic properties over finite 
fields, whereas the present scheme relies on algebraic properties over the 
field $\mathbb{Q}(\alpha)$. For example, 
Pollard's method ends up to build a polynomial $x^n-1$ with $p-1$ roots over a finite
$\mathbb{Z}_p$, where $p$ is some prime factor of $c$. This feature of the polynomial 
comes from Fermat's little theorem and is satisfied if the integer $n$ has $p-1$ as factor.
Thus, the existence of a large number of zeros of $x^n-1$ strictly
depends on the field. Indeed, the polynomial does not have more than $2$ roots over the
rationals. In our scheme, the main task is to find rational functions having a 
sufficiently large number of zeros over an algebraic number field. This feature is then
inherited by the functions over finite fields.
Some specific properties of finite fields can eventually be useful, such as the reducibility
of $P_I$ over $\mathbb{Z}_p$. This will be mentioned later.

Let us illustrate the general idea with an example.
Suppose that the input of the factorization problem is $c=p p'$, with $p$ and $p'$ 
prime numbers and $p<p'$. Let the map be a univariate polynomial of the form 
\be\label{Eq1}
P(x)=\prod_{i=1}^{N_P}(x-x_i), 
\ee
where $x_i$ are integer numbers somehow randomly distributed 
in an interval between $1$ and $i_{max}\gg p'$. More generally, $x_i$ can be rational
numbers $n_i/m_i$ with $n_i$ and/or $m_i$ in $\{1,\dots,i_{max}\}$.
When reinterpreted modulo $p$ or modulo $p'$, the numbers
$x_i$ take random values over the finite fields. If $N_P<p$, we expect that the polynomial
has about $N_P$ distinct roots over the finite fields. Thus, the probability that 
$P(x)\mod p=0$ or $P(x)\mod p'=0$ is about $N_P/p$ or $N_P/p'$, respectively, which are the
ratio between the number of zeros and the size of the input space $\Omega$ over the finite fields.
The probability that $P(x)\mod c$ contains a nontrivial factor of $c$ is about 
$\frac{N_P}{p}\left(1-\frac{N_P}{p'}\right)+\frac{N_P}{p'}\left(1-\frac{N_P}{p}\right)$.
Thus, if $N_P$ is of the order of $p$, we can get a nontrivial factor
by the Euclidean algorithm in few trials. More specifically, if $p\simeq p'$ and 
$N_P\simeq \sqrt{c}/2$, then the probability of getting a nontrivial factor is roughly
$1/2$. It is clear that a computational complexity of the map scaling subexponentially or 
polynomially in $\log N_P$ leads to a subexponential or polynomial
complexity of the factoring algorithm. Thus, the central problem is to build a polynomial
$P(x)$ or a rational function with friendly computational properties with respect to
the number of zeros.
The scheme can be generalized by taking multivariate
maps with $M$ input parameters. In this case, the number of zeros needs to be of the order of 
$p^M$, which is the size of the input space over the field $\mathbb{Z}_p$. As a further 
generalization, the rational field can be replaced by an algebraic number field 
$\mathbb{Q}(\alpha)$ of degree $k_0$. A number in this field is represented as a
$k_0$-dimensional vector over the rationals. Reinterpreting the components of 
the vector over a finite field $\mathbb{Z}_p$, the size of the sampling space
is $p^{k_0 M}$, so that we should have $N_P\sim p^{k_0 M}$ in order to get nontrivial
factors in few trials. Actually, this is the worst-case scenario since the reinterpretation
of $\mathbb{Q}(\alpha)$ modulo $p$ can lead to a degree of the finite field much smaller 
than $k_0$. For example, if $\alpha$ is the root $e^{2\pi i/n}$ of the polynomial $x^n-1$, 
the degree of the corresponding finite field with characteristic $p$ collapses to $1$ if 
$n$ is a divisor of $p-1$.

On one hand, it is trivial to build polynomials with an arbitrarily large number $N_P$
of roots over the rationals as long as the computational cost grows linearly in $N_P$.
On the other hand, it is also simple to build polynomials with friendly computational 
complexity with respect to $\log N_P$ if the roots are taken 
over algebraically closed fields. The simplest example is the previously mentioned
polynomial $P(x)=x^n-1$, which has $n$ distinct complex roots and a computational complexity
scaling as $\log n$. However, over the rationals, this  polynomial has at most 
$2$ roots. We can include other roots by extending the rational field to an algebraic number 
field, but the extension would have a degree proportional to the number of roots, so that
the computational complexity of evaluating $P(x)$ would grow polynomially in the number
of roots over the extension. 
\subsection{Algebraic-geometry rephrasing of the problem}
It is clear that an explicit definition of each root of polynomial~(\ref{Eq1}) leads
to an amount of memory allocation growing exponentially in $\log c$, so that the resulting
factoring algorithm is exponential in time and space. Thus, the roots has to be defined
implicitly by some simple rules. Considering a purely algebraic definition, we associates 
the roots to rational solutions of a set of $n$ non-linear polynomial equations in 
$n$ variables ${\bf x}=(x_1,\dots,x_n)$,
\be
P_k({\bf x})=0,\;\;\;\; k\in\{0,\dots,n-1\}.
\ee
The solutions are intersection points of $n$ hypersurfaces. 
The roots of $P(x)$ are defined as the values of some coordinate, say $x_n$, at the
intersection points. By eliminating the $n-1$ variables $x_1,\dots,x_{n-1}$, we end
up with a polynomial $P(x_n)$ with a number of roots generally growing exponentially in $n$.
This solves the problem of space complexity in the
definition of $P(x)$. There are two remaining problems. First, we have to choose
the polynomials $P_0,\dots,P_{n-1}$ such that an exponentially large fraction of
the intersection points are rational. Second, the variable elimination, given a value 
of $x_n$, has to be performed as efficiently as possible over a finite field. If
the elimination has polynomial complexity, then factorization turns out to have
polynomial complexity. Note that the elimination of $n-1$ variables given $x_n$
is equivalent to a consistency test of the $n$ polynomials. The first problem can be
solved in a simple way by defining the polynomials as elements of the ideal generated
by products of linear polynomials. Let us denote a linear polynomial with a symbol
with a hat. For example, the quadratic polynomials 
\be\label{quadr_polys}
G_i=\hat a_i\hat b_i, \;\;\;\;\ i\in\{1,\dots,n\}
\ee
have generally $2^n$ rational common zeros, provided that the coefficients of $\hat a_i$ 
and $\hat b_i$ are rational.
Identifying the polynomials $P_0,\dots,P_{n-1}$ with elements of the ideal generated
$G_1,\dots,G_n$, we have
\begin{equation}
\label{poly_eqs_simple}
P_k=\sum_i c_{k,i}({\bf x})\hat a_i \hat b_i,\;\;\; i\in\{0,\dots,n-1\},
\end{equation}
whose set of common zeros contains the $2^n$ rational points of the generators $G_i$.
In particular, if the polynomials $c_{k,i}$ are set equal to constants, then
the system  $P_0=\dots P_{n-1}=0$ is equivalent to the system  $G_1=\dots G_n=0$. 

At this point, the variable elimination is the final problem. A working method is to 
compute a Gr\"obner basis. For the purpose of factorizing $c=p p'$, the task is
to evaluate a Gr\"obner basis to check if the $n$ polynomials given $x_n$ are
consistent modulo $c$. If they are consistent modulo some non-trivial factor $p$ of $c$, 
we end up at some point with some integer equal to zero modulo $p$. 
However, the complexity of this computation is doubly exponential in the worst
case. Thus, we have to search for a suitable set of polynomials with a large set of 
rational
zeros such that there is an efficient algorithm for eliminating $n-1$ variables.
The variable elimination is efficient if $n-1$ out of the $n$ polynomial equations 
$P_k=0$ form a suitable triangular system for some set of low-degree polynomials $c_{k,i}$.
Let us assume that the last $n-1$ polynomials have
the triangular form 
$$
\left.
\begin{array}{r}
P_{n-1}(x_{n-1},x_n)  \\
P_{n-2}(x_{n-2},x_{n-1},x_n)  \\
\dots   \\
P_1(x_1,\dots,x_{n-2},x_{n-1},x_n)
\end{array}
\right\},
$$
such that the $k$-th polynomial is linear in $x_k$. Thus, the corresponding polynomial
equations can be sequentially solved in the first $n-1$ variables
through the system
\be
\label{rational_system}
\begin{array}{l}
x_{n-1}=\frac{{\cal N}_{n-1}(x_n)}{{\cal D}_{n-1}(x_n)}  \\
x_{n-2}=\frac{{\cal N}_{n-2}(x_n,x_{n-1})}{{\cal D}_{n-2}(x_n,x_{n-1})}  \\
\dots \\
x_1=\frac{{\cal N}_1(x_n,x_{n-1},\dots,x_2)}{{\cal D}_1(x_n,x_{n-1},\dots,x_2)},
\end{array}
\ee
where ${\cal D}_k\equiv \partial P_k/\partial x_k$ and ${\cal N}_k\equiv P_k|_{x_k=0}$. 
This system defines a parametrizable curve,
say $\cal V$, in the algebraic set defined by the polynomials $P_1,\dots,P_{n-1}$, the variable
$x_n$ being the parameter.
Let us remind that a  curve is parametrizable if and only if its geometric genus is equal to 
zero. 
The overall set of variables can be efficiently computed over a finite field, provided that 
the polynomial coefficients $c_{k,i}$ are not too complex. Once determined the variables,
the remaining polynomial $P_0$ turns out to be equal to zero if ${\bf x}$
is an intersection point. Provided that the rational intersection points have distinct values
of $x_n$ (which is essentially equivalent to state that the points are distinct and 
are in the variety $\cal V$), 
then the procedure generates a value $P_0({\bf x})$ which is zero modulo $p$ with
high probability if $p$ is of the order of the number of rational intersection
points. For this inference, it is pivotal to assume that a large fraction of rational points 
remain distinct when reinterpreted over the finite field.

This algebraic-geometry rephrasing of the problem can be stated in a more general form.
Let $\cal V$ and $\cal H$ be some irreducible curve in an $n$-dimensional space and a hypersurface,
respectively. The curve $\cal V$ is not necessarily parametrizable, thus its genus may take
strictly positive values.  The points in $\cal H$ are the zero locus 
of some polynomial $P_0$. Let $N_P$ be the
number of distinct intersection points between $\cal V$ and $\cal H$ over the rational field $\mathbb{Q}$.
Over a finite field $\text{GF}(q)$, Weil's theorem states that the number of rational
points, says $N_1$, of a smooth curve is bounded by the inequalities
\be\label{weil_bounds}
-2 g \sqrt{q} \le N_1-(q+1)\le 2 g\sqrt{q},
\ee
where $g$ is the geometric genus of the curve.
Generalizing to singular curves~\cite{aubry}, we have
\be\label{aubry_eq}
-2 g \sqrt{q}-\delta \le N_1-(q+1)\le 2 g\sqrt{q}+\delta,
\ee
where $\delta$ is the number of singularities, properly counted. These inequalities have the 
following geometric interpretation. For the sake of simplicity, let us assume that the
singularities are ordinary double points.
A singular curve, says $\cal S$, with genus $g$  is birationally 
equivalent to a smooth curve, say $\cal R$, with same genus, for which Wiel's theorem holds. 
That is, the rational points of $\cal R$ are bijectively mapped to the non-singular
rational points of $\cal S$,
apart from a possible finite set $\Omega$ of $2m$ points mapping to $m$ singular points
of $\cal S$. 
The cardinality of $\Omega$ is at most $2\delta$ (attained when the $\delta$ singularities
have tangent vectors over the finite field). We have two extremal cases. 
In one case, $\#\Omega=2\delta$, so that, $\cal S$ has $\delta$ points less than $\cal R$ 
(two points in $\Omega$ are merged into a singularity of $\cal S$). This gives the lower
bound in~(\ref{aubry_eq}). In the second case, $\Omega$ is empty and the singular points of
$\cal S$ are rational points. Thus, $\cal S$ has $\delta$ rational points more.
Given this interpretation, Weil's upper bound still holds for the number of 
non-singular rational points, say $N_1'$,
\be\label{weil2}
N_1'\le (q+1)+2 g \sqrt{q}.
\ee
Thus, if the genus is much smaller than $\sqrt{q}$, $N_1'$ is upper-bounded by a number close
to the order of the field.

Now, let us assume that most of the $N_P$ rational points in ${\cal V}\cap{\cal H}$ over $\mathbb{Q}$
remain distinct when reinterpreted over $\mathbb{Z}_p$, with $p\simeq a N_P$, where $a$ is a number
slightly greater than $1$, say $a=2$. We also assume that
these points are not singularities of $\cal V$. Weil's inequality~(\ref{weil2}) implies that
the curve does not have more than about $p$ points over $\mathbb{Z}_p$. Since 
$p\gtrsim N_1'\gtrsim N_P\sim p/2$, we have that the number of non-singular rational points of 
the curve is about the number of intersection points over $\mathbb{Z}_p$. This implies that a 
large fraction of points $\bf x\in \cal V$ over the finite field are also points of $\cal H$. We 
have the following.
\begin{claim}
\label{claim_rat_curve}
Let ${\cal V}$ and ${\cal H}$ be an algebraic curve with genus $g$ and a hypersurface, respectively.
The hypersurface is the zero locus of the polynomial $P_0$.
Their intersection has $N_P$ distinct points over the rationals, which are not singularities
of $\cal V$. Let us also assume that $g\ll \sqrt{N_P}$ and that most of the $N_P$ rational points remain 
distinct over $\mathbb{Z}_p$ with $p\gtrsim 2 N_P$. 
If we pick up at random a point in $\cal V$, then 
\be
P_0({\bf x})=0 \mod p
\ee
with probability equal to about ratio $N_P/p$.
\end{claim}
If there are pairs $({\cal V},{\cal H})$ for every $N_P$ that satisfy the premises of this claim,
then prime factorization is reduced to the search of rational points of a curve. Actually,
these pairs always exist, as shown later in the section. Assuming that
$c=p p'$ with $p\sim p'$, the procedure for factorizing $c$ is as follows.
\begin{enumerate}
\item[(1)] Take a pair $({\cal V},{\cal H})$ with $N_P\sim c^{1/2}$ such that the premises of
Claim~\ref{claim_rat_curve} hold.
\item[(2)] Search for a rational point ${\bf x}\in\cal V$ over $Z_p$.
\item[(3)] Compute $\text{GCD}[P_0({\bf x}),c]$, the greatest common divisor of $P_0({\bf x})$ and $c$.
\end{enumerate}
The last step gives $\text{GCD}[P({\bf x}),c]$ equal to 
$1$, $c$ or one of the factors of $c$. The probability of getting
a nontrivial factor can be made close to $1/2$ with a suitable tuning of $N_P$ (as shown later in 
Sec.~\ref{sec_algo}).

Finding rational points of a general curve with genus greater than $2$ is an exceptionally complex 
problem. For example, just to prove that the plane curve $x^h+y^h=1$ with $h>2$ does not have zeros over 
the rationals took more than three centuries since Fermat stated it.
Curves with genus $1$ are elliptic curves, which have an important role
in prime factorization (See Lenstra algorithm~\cite{lenstra}). Here, we will focus on
parametrizable curves, which have genus $0$. In particular, we will consider parametrizations
generated by the sequential equations~(\ref{rational_system}). It is interesting to note that
it is always possible to find a curve $\cal V$ with parametrization~(\ref{rational_system})
and a hypersurface $\cal H$ such that their intersection contains a given set of rational points.
In particular, there is a set of polynomials $P_0,\dots,P_{n-1}$ of the form~(\ref{poly_eqs_simple}),
such that the zero locus of the last $n-1$ polynomials contains a parametrizable curve with
parametrization~(\ref{rational_system}), whose intersection with the hypersurface $P_0=0$ contains
$2^n$ distinct rational points. Let the intersection points be defined by the 
polynomials~(\ref{quadr_polys}). Provided that $x_n$ is a separating variable, the set of
intersection points admits the rational univariate representation~\cite{rouillier}
\be
\left\{
\begin{array}{l}
x_{n-1}=\frac{\bar{\cal N}_{n-1}(x_n)}{{\bar{\cal D}}_{n-1}(x_n)}  \\
x_{n-2}=\frac{{\bar{\cal N}}_{n-2}(x_n)}{\bar{\cal D}_{n-2}(x_n)}  \\
\dots \\
x_1=\frac{\bar{\cal N}_1(x_n)}{\bar{\cal D}_1(x_n)}   \\
{\bar{\cal N}}_0(x_n)=0 
\end{array}
\right.
\ee
The first $n-1$ equations are a particular form of equation~(\ref{rational_system}) and
define a parametrizable curve with $x_n$ as parameter. The last equation can be replaced
by some linear combination of the polynomials $G_i$. It is also interesting
to note that the rational univariate representation is unique once the separating
variable is chosen. This means that the parametrizable curve is uniquely determined by
the set of intersection points and the variable that is chosen as parameter. 

It is clear that the curve and hypersurface obtained through this construction with a 
general set of polynomials $G_i$ satisfy
the premises of Claim~\ref{claim_rat_curve}. Indeed, a large part of the common zeros 
of the polynomials $G_i$ are generally distinct over a finite field $\mathbb{Z}_p$
with $p\simeq N_P$. For example, the point $\hat a_1=\dots=\hat a_n=0$ is distinct from the other 
points if and only if $\hat b_i\ne 0$ at that point for every $i\in\{1,\dots,n\}$.
Thus, the probability that a given point is not distinct over $\mathbb{Z}_p$ is of 
the order of $p^{-1}\sim N_P^{-1}$, hence a large part of the points are distinct over the finite
field. There is an apparent paradox. With a suitable choice of the linear functions $\hat a_i$ 
and $\hat b_i$, the intersection points can be made distinct over a field $\mathbb{Z}_p$ with
$p\ll N_P$, which contradicts Weil's inequality~(\ref{weil2}). The contradiction is explained
by the fact that the curve is broken into the union of reducible curves over the finite field.
In other words, some denominator ${\cal D}_k$ turns out to be equal to zero modulo $p$
at some intersection points. This may happen also with $p\sim N_P$, which is not a concern.
Indeed, possible zero denominators can be used to find factors of $c$.

\subsection{Contents}

In Section~\ref{sec_algo}, we introduce the general scheme of the factoring algorithm
based on rational maps and discuss
its computational complexity in terms of the complexity of the maps, the number
of parameters and the field degree. In Section~\ref{sec_arg_geo}, the factorization 
problem is reduced to the search of rational points of parametrizable algebraic varieties
$\cal V$ having an arbitrarily large number $N_P$ of rational intersection points
with a hypersurface $\cal H$. Provided that the $N_P$ grows exponentially in the
space dimension, the factorization algorithm has polynomial complexity if the
number of parameters and the complexity of evaluating a point in $\cal V$  over a 
finite field grow sublinearly and polynomially in the space dimension, respectively.
Thus,
the varieties $\cal V$ and $\cal H$ have to satisfy two requirements.
On one side, their intersection has to contain a large set of rational points.
On the other side, $\cal V$ has to be parametrizable and its points 
have to be computed efficiently given
the values of the parameters. The first requirement is fulfilled with a generalization
of the construction given by Eq.~(\ref{poly_eqs_simple}). First, we define an ideal
$I$ generated by products of linear polynomials such that the associated algebraic
set contains $N_P$ rational points. The relevant information on this ideal
is encoded in a satisfiability formula (SAT) in conjunctive normal form (CNF) and a 
linear matroid. Then, we define $\cal V$ and $\cal H$ as elements of the ideal.
By construction, ${\cal V}\cap{\cal H}$ contains the $N_P$ rational points.
The ideal $I$ and the polynomials defining the varieties contain some coefficients.
The second requirement is tackled in Sec.~\ref{build_up}. By imposing the 
parametrization of $\cal V$, we get a set of polynomial equations for the
coefficients. These equations always admit a solution, provided that the
only constraint on $\cal V$ and $\cal H$ is being an element of $I$.
The task is to find
an ideal $I$ and a set of coefficients such that the computation of points
in $\cal V$ is as efficient as possible, given a number of parameters scaling
sublinearly in the space dimension.

In this general form, the problem of building the varieties $\cal V$ and
$\cal H$ is quite intricate. A good strategy is to start with simple ideals
and search for varieties in a subset of these ideals, so that 
the polynomial constraints on the unknown coefficients can be handled with
little efforts. With these restrictions, it is not guaranteed that 
the required varieties exist, but we can have hints on how to proceed.
This strategy is employed in Sec.~\ref{sec_quadr_poly}, where we consider an 
ideal generated by the polynomials~(\ref{quadr_polys}). The varieties
are defined by linear combinations of these generators with constant
coefficients, that is, $\cal H$ and $\cal V$ are in the zero locus of 
$P_0$ and $P_1,\dots,P_{n-1}$, respectively, defined
by Eq.~(\ref{poly_eqs_simple}). The $2^n$ rational points associated
with the ideal are taken distinct in $\mathbb{Q}$. First, we prove 
that there is no solution with one parameter ($M=1$), for a dimension
greater than $4$. We give an explicit numerical example of a curve
and hypersurface in dimension $4$. The intersection has $16$ rational
points. We also give a solution with about $n/2$ parameters. Suggestively,
this solution resembles a kind of retro-causal model. Retro-causality
is considered one possible explanation of some strange aspects of
quantum theory, such as non-locality and wave-function collapse after
a measurement. Finally, we close the section by proving that there is
a solution with $2\le M \le (n-1)/3$. This is shown by explicitly
building a variety $\cal V$ with $(n-1)/3$ parameters. Whether
it is possible to drop the number of parameters below this upper
bound is left as an open problem. If $M$ grows
sublinearly in $n$, then there is automatically a factoring
algorithm with polynomial complexity, provided that the coefficients
defining the polynomials $P_k$ are in $\mathbb{Q}$ and can be
computed efficiently over a finite field. The conclusions and 
perspectives are drawn in Sec.~\ref{conclusion}.

\section{General scheme and complexity analysis}
\label{sec_algo}
At a low level, the central object of the factoring algorithm under study is a
class $\Theta$ of maps ${\vec\tau}\mapsto {\cal R}(\vec\tau)$ 
from  a set $\vec\tau\equiv(\tau_1,\dots,\tau_M)$ of $M$ parameters over the field $\mathbb{Q}(\alpha)$ to a 
number in the same field, where $\cal R$ is a rational function, that is, the algebraic fraction of 
two polynomials. Let us write it as
$$
{\cal R}(\vec\tau)\equiv \frac{{\cal N}(\vec\tau)}{{\cal D}(\vec\tau)}.
$$
This function may be indirectly defined by applying consecutively simpler rational functions, as 
done in Sec.~\ref{sec_arg_geo}. Note that the computational complexity of evaluating ${\cal R}(\vec\tau)$ 
can be lower than the complexity of evaluating the numerator ${\cal N}(\vec\tau)$. For this reason we 
consider more general rational functions rather than polynomials.
Both $M$ and $\alpha$ are not necessarily fixed in the class $\Theta$.
We denote by $N_P$ the number of zeros of the polynomial $\cal N$ over  $\mathbb{Q}(\alpha)$. 
The number $N_P$ is supposed to be finite, we will come back to this assumption later in
Sec.~\ref{sec_infinite_points}.
For the sake of simplicity, first we introduce the general scheme of the algorithm over the 
rational field. Then, we outline its extension to algebraic number fields.
We mainly consider the case of semiprime input, that is, $c$ is
taken equal to the product of two prime numbers $p$ and $p'$. This case is the most
relevant in cryptography. If the rational points are somehow randomly distributed
when reinterpreted over $\mathbb{Z}_p$, then the polynomial ${\cal N}$ has at least about 
$N_P$ distinct zeros over the finite field, provided that $N_P$ is sufficiently smaller than 
the size $p^{M}$ of the input space $\Omega$.
We could have additional zeros in the finite field, but we conservatively assume that
$N_P$ is a good estimate for the total number.
For $N_P$ close to $p^M$, two different roots in the rational field may collapse to the same 
number in the finite field. We will account for that later in Sec.~\ref{sec_complex}.
Given the class $\Theta$, the factorization procedure has the same general scheme as other methods
using finite fields. Again, the value $m={\cal R}(\tau_1,\tau_2,\dots)$ is computed by pretending
that $c$ is prime and $\mathbb{Z}/c\mathbb{Z}$ is a field. If an algebraic division takes to a 
contradiction during the computation of $\cal R$, the divisor is taken as outcome $m$. For the sake of 
simplicity, we neglect the zeros of $\cal D$ and consider only the zeros of $\cal N$.
In Section~\ref{sec_arg_geo}, we will see that this 
simplification is irrelevant for a complexity analysis. It is clear that the outcome $m$ is zero 
in $\mathbb{Z}/c\mathbb{Z}$ with high probability for some divisor $p$ of $c$ if the number of 
zeros is about or greater than the number of inputs $p^{M}$. Furthermore, if $p'>p$ and 
$N_P$ is sufficiently smaller than $(p')^{M}$, then the outcome $m$
contains the nontrivial factor $p$ of $c$ with high probability. This is guaranteed if 
$N_P$ is taken equal to about $c^{M/2}$, which is almost optimal if $p\simeq p'$, as we will see later
in Sec.~\ref{sec_complex}. Thus, we have the following.
\begin{algorithm} 
\label{gen_algo0}
Factoring algorithm with input $c=p p'$, $p$ and $p'$ being prime numbers.
\item[(1)] \label{algo0_1}
Choose a map in $\Theta$ with $M$ input parameters and $N_P$ zeros over the rationals
such that $N_P\simeq c^{M/2}$ (see Sec.~\ref{sec_complex} for an optimal choice of $N_P$);
\item[(2)] generate a set of $M$ random numbers $\tau_1,\dots,\tau_M$ over $\mathbb{Z}/c\mathbb{Z}$.
\item[(3)] \label{algo0_3}
compute the value $m={\cal R}(\tau_1,\dots,\tau_M)$ over $\mathbb{Z}/c\mathbb{Z}$
(by pretending that $c$ is prime). 
\item[(4)] \label{algo0_4}
compute the greatest common divisor between $m$ and $c$.
\item[(5)] if a nontrivial factor of $c$ is not obtained, repeat from point (2).
\end{algorithm}
The number $M$ of parameters may depend on the map picked 
up in $\Theta$. Let $M_{min}(N_P)$ be the minimum of $M$ in $\Theta$ for given $N_P$. 
The setting at point~(\ref{algo0_1}) is possible only if $M_{min}$ grows less than linearly in
$\log N_P$, which is condition~(c) enumerated in the introduction. A tighter condition is
necessary if the computational complexity of evaluating the map scales subexponentially, but
not polynomially. This will be discussed with more details in Sec.~\ref{sec_complex}.

If $c$ has more than two prime factors,
$N_P$ must be chosen about equal to about $p^{M}$, where $p$ is an estimate of one
prime factor.
If there is no knowledge about the factors, the algorithm can be executed by trying different 
orders of magnitude of $N_P$ from $2$ to $c^{1/2}$. For example, we can 
increase the guessed $N_P$ by a factor $2$, so that the overall number of executions grows
polynomially in $\log_2 p$. However, better strategies are available. 
A map with a too great $N_P$ ends up to produce zero modulo $p$ for every factor $p$ of $c$
and, thus, the algorithm always generates the trivial factor $c$. Conversely, a too small $N_P$
gives a too small probability of getting a factor. Thus, we can employ a kind of bisection 
search. A sketch of the search algorithm is as follows.
\begin{enumerate}
\item set $a_d=1$ and $a_u=c^M$;
\item set $N_P=\sqrt{a_d a_u}$ and choose a map in $\Theta$ with $N_P$ zeros;
\item execute Algorithm~\ref{gen_algo0} from point (2) and break the loop after a certain number
of iterations;
\item if a nontrivial factor is found, return it as outcome;
\item if the algorithm found only the trivial divisor $c$, set $a_u=N_P$, otherwise set $a_d=N_P$;
\item go back to point (2).
\end{enumerate}
This kind of search can 
reduce the number of executions of Algorithm~\ref{gen_algo0}. In the following, we will not
discuss these optimizations for multiple prime factors, we will consider mainly semiprime integer 
numbers $c=p p'$.

\subsection{Extension to algebraic number fields}

Before outlining how the algorithm can be extended to algebraic number fields,
let us briefly remind what a number field is.
The number field $\mathbb{Q}(\alpha)$ is a rational field extension obtained by adding 
an algebraic number $\alpha$ to the field $\mathbb{Q}$. 
The number $\alpha$ is solution of some irreducible polynomial $P_I$ of degree $k_0$, which is also
called the degree of $\mathbb{Q}(\alpha)$. The extension 
field includes all the elements of the form $\sum_{i=0}^{k_0-1} r_i \alpha^i$, where $r_i$ are rational
numbers. Every power $\alpha^h$ with $h\ge k_0$ can be reduced to that form through the equation
$P_I(\alpha)=0$.
Thus, an element of $\mathbb{Q}(\alpha)$ can be represented as a $k_0$-dimensional vector over
$\mathbb{Q}$. Formally, the extension field is defined as the quotient ring $\mathbb{Q}[X]/P_I$,
the polynomial ring over $\mathbb{Q}$ modulo $P_I$. The quotient ring is also a field as long as
$P_I$ is irreducible.
Reinterpreting the rational function $\cal R$ over a finite field $\text{GF}(p^k)$ means to reinterpret 
$r_i$ and the coefficients of $P_I$ as integers modulo a prime number $p$. 
Since the polynomial $P_I$ may be reducible over $\mathbb{Z}_p$, the degree $k$ of
the finite field is some value between $1$ and $k_0$ and equal to the degree of one the
irreducible factors of $P_I$. Let $D_1,\dots,D_f$ be the factors of $P_I$. Each $D_i$ is
associated with a finite field $\mathbb{Z}_p[X]/D_i\cong \text{GF}(p^{k_i})$, where $k_i$ is
the degree of $D_i$. Smaller values of $k$ take to a computational advantage, as the
size $p^{k M}$ of the input space $\Omega$ is smaller and the probability, about $N_P/p^{k M}$, of
getting the factor $p$ is higher. For example, the cyclotomic number field with $\alpha=e^{2\pi i/n}$
has a degree equal to $\phi(n)$, where $\phi$ is the Euler totient function, which is asymptotically
lower-bounded by $K n/\log\log n$, for every constant $K<e^{-\gamma}$, $\gamma$ being the Euler
constant. In other words, the highest degree of the polynomial prime factors of $x^n-1$ is equal to
$\phi(n)$. Let $P_I$ be equal to the factor with $e^{2\pi i/n}$ as root.
If $n$ is a divisor of $p-1$ for some prime number $p$, then $P_I$ turns out to have linear
factors over $\mathbb{Z}_p$. Thus, the degree of the number field collapses to $1$
when mapped to a finite field with characteristic $p$. Thus, the bound $k_0$ sets a worst case. 

For general number fields, the equality $k=k_0$ is more an exception than a rule, apart from the 
case of the rational field, for which $k=k_0=1$. For the sake of simplicity, let us assume
for the moment that $k=k_0$ for one of the two factors of $c$, say $p'$. Algorithm~\ref{gen_algo0} 
is modified as follows. The map is chosen at point~(1) of Algorithm~\ref{gen_algo0} such that 
$N_P\simeq c^{k_0 M/2}$;
the value $m$ computed at point~(3) is a polynomial over $\mathbb{Z}/c\mathbb{Z}$ of 
degree $k_0-1$; the greatest common divisor at point~(4) is computed
between one of the coefficients of the polynomial $m$ and $c$. If the degree $k$ of the finite
field of characteristic $p$ turns out to be smaller than $k_0$, we have to compute the 
polynomial greatest common divisor between $m$ and $P_I$ by pretending again that 
$\mathbb{Z}/c\mathbb{Z}$ is a field. If $m$ is zero over $\text{GF}(p^{k})$, 
then the Euclidean algorithm generates at some point 
a residual polynomial with the leading coefficient having $p$ as a factor (generally, all the
coefficients turn out to have $p$ as a factor).
If $k\ne k_0$ for both factors and most of the maps, then the algorithm ends up to generate
the trivial factor $c$, so that we need to decrease $N_P$ until a non-trivial factor is
found.

\subsection{Complexity analysis}
\label{sec_complex}

The computational cost of the algorithm grows linearly with the product between the computational 
cost of the map, say ${\bf C}_0({\cal R})$, and the average number of trials, which is
roughly $p^{k_0 M}/N_P$ provided that $N_P\ll p^{k_0 M}$ and $P_I$ is irreducible over 
$\mathbb{Z}_p$. The class $\Theta$ may contain many maps with a given number $N_P$ of
zeros over some number field. We can choose the optimal map for each $N_P$, so that we express 
$k_0$, $M$ and $\cal R$ as functions of $\log N_P\equiv \xi$. The computational cost
${\bf C}_0({\cal R})$ is written as a function of $\xi$, ${\bf C}_0(\xi)$.

Let us evaluate the computational complexity of the algorithm in terms of the scaling
properties of $k_0(\xi)$, $M(\xi)$ and ${\bf C}_0({\xi})$ as functions of $\xi=\log N_P$.
The complexity ${\bf C}_0(\xi)$ is expected to be a monotonically increasing
function. If the functions $k_0(\xi)$ and $M(\xi)$ were decreasing,
then they would asymptotically tend to a constant, since they are not less than  $1$.
Thus, we assume that these two functions are monotonically increasing or constant.

As previously said, the polynomial $\cal N$ has typically about $N_P$ distinct roots
over $\text{GF}(p^{k_0})$, provided that $N_P$ is sufficiently smaller than $p^{k_0 M}$. 
If $N_P$ is greater than $p^{k_0 M}$, then almost every
value of $\vec\tau$ is a zero of the polynomial. Assuming that the zeros are somehow
randomly distributed, the probability that a number picked up at random is different from
any zero over $\text{GF}(p^{k_0})$ is equal to $(1-p^{-k_0 M})^{N_P}$.
Thus, the number of roots over $\text{GF}(p^{k_0})$ is expected to be of the order of
$p^{k_0 M} [1-(1-p^{-k_0 M})^{N_P}]$, which is about $N_P$ for $N_P\ll p^{k_0 M}$. Thus,
the average number of trials required for getting a zero is
\be
N_{trials}\equiv \frac{1}{1-(1-p^{-k_0 M})^{N_P}},
\ee
A trial is successful if it gives
zero modulo some nontrivial factor of $c$, thus the number of required trials can
be greater than $N_{trials}$ if some factors are close each other. Let us consider the 
worst case with $c=p p'$, where $p$ and $p'$ are two primes with $p'\simeq p$ such
that $(p')^{k_0 M}\simeq p^{k_0 M}$. 
Assuming again that the roots are randomly distributed, the probability of a successful
trial is $\text{Pr}_\text{succ}\equiv 2 [1-(1-p^{-k_0 M})^{N_P}](1-p^{-{k_0} M})^{N_P}$. 
The probability has a maximum equal to $1/2$ for $\xi$ equal to the value 
\be
\xi_0\equiv\log\left[-\frac{\log 2}{\log(1-p^{-k_0 M})}\right].
\ee
Evaluating the Taylor series at $p=\infty$, we have that
\be
\xi_0=k_0 M\log p+\log\log 2-\frac{1}{2 p^{k_0 M}}+O(p^{-2 k_0 M}).
\ee
The first two terms give a very good approximation of $\xi_0$.
At the maximum, the ratio between the number of zeros and the number of states $p^{k_0 M}$ 
of the sampling space is about $\log 2$. It is worth to note that,
for the same value of $\xi$, the probability of getting an isolated factor 
with $p\ll p'$ is again exactly $1/2$. Thus, we have in general
\be
N_P\simeq 0.69 p^{k_0 M} \Rightarrow \text{Pr}_\text{succ}=1/2.
\ee
Since the maximal probability is independent of $k_0$ and $M$, this
value is also maximal if $k_0$ and $M$ are taken as functions of $\xi$. 
The maximal value $\xi_0$ is solution of the equation 
\be
\label{eq_xi0}
\xi_0=\log\left[-\frac{\log 2}{\log(1-p^{-f(\xi_0)})}\right].
\ee
where $f(\xi)\equiv k_0(\xi) M(\xi)$. If the equation has no positive solution,
then the probability is maximal for $\xi=0$. That is, the optimal map in the
considered class is the one with $N_P=1$. This means that the number of states
$p^{k_0 M}$ of the sampling space grows faster than the number of zeros.
In particular, there is no solution for $\log p$ sufficiently large if
$f(\xi)$ grows at least linearly (keep in mind that $f(\xi)\ge1$). Thus,
the function $f(\xi)$ has to be a sublinear power function, as previously 
said.

The computational cost of the algorithm for a given map ${\cal R}(\xi)$ is
\be
{\bf C}(p,\xi)\equiv 
\frac{{\bf C}_0(\xi)}{2 [1-(1-p^{-f(\xi)})^{\exp\xi}](1-p^{-f(\xi)})^{\exp\xi }}.
\ee
The optimal map for given $p$ is obtained by minimizing ${\bf C}(p,\xi)$ with 
respect to $\xi$. The computational complexity of the algorithm is
\be\label{comp_complexity}
{\bf C}(p)=\min_{\xi>0} {\bf C}(p,\xi)\equiv {\bf C}(p,\xi_m),
\ee
which satisfies the bounds
\begin{equation}
\label{bounds}
{\bf C}_0(\xi_m)\le {\bf C}(p)\le 2{\bf C}_0(\xi_0).
\end{equation}
The upper bound in Eq.~(\ref{bounds}) is the value of ${\bf C}(p,\xi)$ at $\xi=\xi_0$,
whereas the lower bound is the computational complexity of the map at the minimum
$\xi_m$.

It is intuitive that the complexity ${\bf C}_0(\xi)$ must be subexponential in order to 
have ${\bf C}(p)$ subexponential in $\log p$. This can be
shown by contradiction. Suppose that the complexity ${\bf C}(p)$ is subexponential
in $\log p$ and ${\bf C}_0(\xi)=\exp(a \xi)$ for some positive $a$. The lower bound
in Eq.~(\ref{bounds}) implies that the optimal $\xi_m$ grows less than $\log p$. 
Asymptotically, 
\be
\left. \frac{p^{k_0 M}}{N_P}\right|_{\xi=\xi_m}\sim e^{f(\xi_m)\log p-\xi_m}\ge K p^{1/2},
\ee
for some constant $K$.
Thus, the average number of trials grows exponentially in $\log p$, implying
that the computational complexity is exponential, in contradiction
with the premise. 
Since $f(\xi)$ and $\log {\bf C}_0(\xi)$ must grow less than linearly, we may assume
that they are concave.
\begin{property} 
\label{concave}
The functions $f(\xi)$ and $\log {\bf C}_0(\xi)$ are concave, that is,
\be
\frac{d^2}{d\xi^2}f(\xi)\le 0, \;\;\;  \frac{d^2}{d\xi^2}\log {\bf C}_0(\xi)\le 0.
\ee
\end{property}

The lower bound in Eq.~(\ref{bounds}) depends on $\xi_m$, which depends on the
function $C_0(\xi)$. A tighter bound which is also simpler to evaluate can be derived
from Property~\ref{concave} and the inequality
\be
{\bf C}(p,\xi)\ge \frac{1}{2}e^{f(\xi)\log p-\xi}{\bf C}_0(\xi).
\ee
\begin{lemma}
If Property~\ref{concave} holds and ${\bf C}(p)$ is asymptotically sublinear in $p$, then there 
is an integer $\bar p$ such that the complexity 
${\bf C}(p)$ is bounded from below by $\frac{{\bf C}_0(\xi_0)}{2\log2}$ for $p>\bar p$.
\end{lemma}
{\it Proof}.
The minimum $\xi_m$ is smaller than $\xi_0$, since the function ${\bf C}_0(\xi)$ is monotonically increasing.
Thus, we have
\be
{\bf C}(p)=\min_{\xi\in\{0,\xi_0\}}{\bf C}(p,\xi)\ge \min_{\xi\in\{0,\xi_0\}}
e^{f(\xi)\log p-\xi+\log C_0(\xi)}/2.
\ee
Since the exponential is monotonic and the exponent is concave, the objective function 
has a maximum and two local minima at the $\xi=0$ and $\xi=\xi_0$. Keeping in mind that $f(\xi)\ge 1$,
The first local minimum is not less than $p {\bf C}_0(0)/2$. The second minimum is 
$e^{f(\xi_0)\log p-\xi_0} {\bf C}_0(\xi_0)/2$,
which is greater than or equal to ${\bf C}_0(\xi_0)/(2 \log2)$. This can be proved by eliminating $p$ through 
Eq.~(\ref{eq_xi0}) and minimizing in $\xi_0$. Since  ${\bf C}(p)$ is sublinear in $p$,
there is an integer $\bar p$ such that the second minimum is global for $p>\bar p$.  $\square$

Summarizing, we have
\begin{equation}
\label{bounds2}
0.72{\bf C}_0(\xi_0)\le {\bf C}(p)\le 2{\bf C}_0(\xi_0)
\end{equation}
for $p$ greater than some integer.
Thus, the complexity analysis of the algorithm is reduced to study the asymptotic behavior
of ${\bf C}_0(\xi_0)$. 
The upper bound is asymptotically tight, that is, $\xi=\xi_0$ is asymptotically optimal. Taking 
$$
f(\xi)=b \xi^\beta \text{   with   }  \beta\in[0:1),
$$
the optimal value of $\xi$ is
$$
\xi_0=(b \log p)^\frac{1}{1-\beta}+O(1).
$$
The function $f(\xi)$ cannot be linear, but we can take it very close to a linear function,
\be
f(\xi)=b \frac{\xi}{(\log\xi)^\beta}, \;\;\;\;  \gamma>1.
\ee
In this case, the optimal $\xi$ is
$$
\xi_0=e^{(b\log p)^{1/\beta}}+O\left[(\log p)^{1/\beta} \right].
$$
There are three scenarios taking to subexponential or polynomial complexity.
\begin{enumerate}
\item[(a)] The functions ${\bf C}_0(\xi)$ and $f(\xi)$ scale polynomially as $\xi^\alpha$ and 
$\xi^\beta$, respectively, with $\beta\in[0:1)$. Then, the computational complexity ${\bf C}(p)$
scales polynomially in $\log p$ as $(\log p)^\frac{\alpha}{1-\beta}$.
\item[(b)] The function ${\bf C}_0(\xi)$ is polynomial and $f(\xi)\sim\xi/(\log\xi)^\beta$ with $\beta>1$. 
Then the computational complexity ${\bf C}(p)$ scales subexponentially in $\log p$ as 
$\exp\left[b (\log p)^{1/\beta}\right]$.
\item[(c)] The function ${\bf C}_0(\xi)$ and $f(\xi)$ are superpolynomial and polynomial respectively,
with ${\bf C}_0(\xi)\sim\exp\left[b \xi^\alpha\right]$ and $f(\xi)\sim\xi^\beta$. If 
$\alpha+\beta<1$, then the complexity ${\bf C}(p)$ is subexponential in $\log p$ and scales
as $\exp\left[b (\log p)^\frac{\alpha}{1-\beta}\right]$.
\end{enumerate}
The algorithm has polynomial complexity in the first scenario. The other cases are subexponential.
This is also implied by the following.
\begin{lemma}
\label{litmus}
The computational complexity ${\bf C}(p)$ is subexponential or polynomial in $\log p$ if 
the function ${\bf C}_0(\xi)^{f(\xi)}$ grows less than exponentially, that is, if 
$$
\lim_{\xi\rightarrow\infty}\frac{f(\xi)\log {\bf C}_0(\xi)}{\xi}=0.
$$
In particular, the complexity is polynomial if ${\bf C}_0(\xi)$ is polynomial
and $f(\xi)$ scales sublinearly.
\end{lemma}
This lemma can be easily proved directly from Eq.~(\ref{eq_xi0}) and the upper bound in 
Eq.~(\ref{bounds}), the former implying the inequality $\xi_0\le f(\xi_0)\log p+\log\log 2$.
Let us prove the first statement. 
$$
\lim_{p\rightarrow\infty}\frac{\log{\bf C}(p)}{\log p}\le
\lim_{\xi\rightarrow\infty}\frac{f(\xi)\log 2 {\bf C}_0(\xi)}{\xi-\log\log 2}=
\lim_{\xi\rightarrow\infty}\frac{f(\xi)\log{\bf C}_0(\xi)}{\xi}=0.
$$
Using the lower bound in Eq.~(\ref{bounds2}),
the lemma can be strengthened by adding the inferences in the other directions 
(\emph{if} replaced by {\emph{if and only if}). 

Summarizing, we have the following.
\begin{claim}
\label{claim1}
The factoring algorithm~\ref{gen_algo0} has subexponential (\emph{polynomial}) complexity if,
for every $\xi=\log N_P>0$ with $N_P$ positive integer, there are rational univariate functions
${\cal R}(\vec\tau)=\frac{{\cal N}(\vec\tau)}{{\cal D}(\vec\tau)}$ of 
the parameters $\vec\tau=(\tau_1,\dots,\tau_{M(\xi)})$ over
an algebraic number field $\mathbb{Q}(\alpha)$ of degree $k_0(\xi)$
with polynomials $\cal N$ and $\cal D$ coprime, such that
\begin{enumerate}
\item the number of distinct roots of $\cal N$ in $\mathbb{Q}(\alpha)$ is equal to
about $N_P$. Most of the roots remain distinct when interpreted over finite fields 
of order equal to about $N_P^{1/M}$;
\item given any value $\vec\tau$, the computation of ${\cal R}(\vec\tau)$ 
takes a number ${\bf C}_0(\xi)$ of arithmetic operations growing less than
exponentially (\emph{polynomially}) in $\xi$;
\item the function ${\bf C}_0(\xi)^{k_0(\xi) M(\xi)}$ is subexponential (\emph{the function 
$k_0(\xi) M(\xi)$ scales sublinearly}).
\end{enumerate}
\end{claim}
Let us stress that the asymptotic complexity is less than exponential if and only if
${\bf C}_0(\xi)^{f(\xi)}$ is less than exponential. Thus, the latter 
condition is a litmus test for a given class of rational functions. However,
the function ${\bf C}_0(\xi)^{f(\xi)}$ does not provide sufficient information on 
the asymptotic computational complexity of the factoring algorithm.
The general number-field sieve is the algorithm
with the best asymptotic complexity, which scales as $e^{a (\log p)^{1/3}}$.
Thus, algorithm~\ref{gen_algo0} is asymptotically more efficient than the general number-field
sieve if ${\bf C}_0(\xi)$ and $f(\xi)$ are asymptotically upper-bounded by a subexponential function
$e^{b(\log p)^\alpha}$ and a power function $c \xi^\beta$, respectively, such that $\alpha<(1-\beta)/3$.
In the limit case of $\beta\rightarrow 1$ and polynomial complexity of the map, the
function $f(\xi)$ must be asymptotically upper-bounded by $b \xi/(\log\xi)^3$.

\subsection{Number of rational zeros versus polynomial degree}
Previously we have set upper bounds on the required computational complexity of the 
rational function $\cal R={\cal N}/{\cal D}$ in terms of the number of its rational zeros.
For a polynomial (subexponential) complexity of prime factorization, the computational complexity 
${\bf C}_0$ of $\cal R$ must scale polynomially (subexponentially) in the logarithm in the number 
of rational zeros. Thus, for a univariate rational function, it is clear that ${\bf C}_0$ 
has to scale polynomially (subexponentially) in the logarithm of the degree $d$ of $\cal N$,
since the number of rational zeros is upper-bounded by the degree 
(fundamental theorem of algebra). An extension of this inference to multivariate 
functions is more elaborate, as upper bounds on the number of rational zeros
are unknown. However, we are interested more properly to a set of $N_P$ rational zeros 
that remain in great part distinct when reinterpreted over a finite field whose order 
is greater than about
$N_P^{1/M}$.
Under this restriction, let us show that the number of rational
zeros of a polynomial of degree $d$ and with $M$ variables is upper-bounded by
$K d^{2 M}$ with some constant $K>0$.
This bound allows us to extend the previous inference on ${\bf C}_0$ to the case of 
multivariate functions.

Assuming that the $N_P$ rational zeros over $\mathbb{Q}$ are randomly distributed
when reinterpreted over $\text{GF}(q)$, their number over the finite field is 
about $q^{M}\left[1-(1-q^{-M})^{N_P}\right]$, as shown previously. Since
an upper bound on the number of zeros $N(q)$ of a smooth hypersurface
over a finite field of order $q$  is known, we can evaluate an upper bound on $N_P$. 
Given the inequality~\cite{katz}
\be\label{gen_weil}
N(q)\le 
\frac{q^M-1}{q-1} +\left[(d-1)^M-(-1)^M\right]\left(1-d^{-1}\right)q^{(M-1)/2}
\ee
and 
\be
\label{bound_ff}
q^M\left[1-(1-q^{-M})^{N_P}\right]\le N(q),
\ee
we get an upper bound on $N_P$ for each $q$.
Requiring that Eq.~(\ref{bound_ff}) 
is satisfied for every $q>N_P^{1/M}$, we get 
\be\label{up_bound}
N_P< K d^\frac{2 M^2}{M+1}< K d^{2 M}
\ee
for some constant $K$ (the same result is obtained by assuming that Eq.~(\ref{bound_ff}) holds
for every $q$).
Note that a slight break of bound~(\ref{up_bound}) with $N_P$ growing as $d_0^{M^a}$
in $M$ for some particular $d=d_0$ and $a>1$ would make the complexity of prime factorization
polynomial, provided the computational complexity of evaluating the function ${\cal R}$ 
is polynomial in $M$. This latter condition can be actually fulfilled, as shown with an
example later.
Ineq.~(\ref{gen_weil}) holds for smooth irreducible hypersurfaces. However, dropping
these conditions are not expected to affect the bound~(\ref{up_bound}). For example, if
$M=2$, then Ineq.~(\ref{gen_weil}) gives
\be\label{weil_plane}
N_P\le q+1+(d-1)(d-2)\sqrt{q}
\ee
which is the Weil's upper bound~(\ref{weil_bounds}) for
a smooth plane curve, whose geometric genus $g$ is equal to $(d-1)(d-2)/2$. 
This inequality holds also for singular curves~\cite{aubry}. Indeed, this comes 
from the upper bound~(\ref{aubry_eq}) and the equality $g=(d-1)(d-2)/2-\delta$.
Also reducibility does not affect Ineq~(\ref{up_bound}).

It is simple to find examples of multivariate functions with a number of rational
zero quite close to the bound $K d^{2 M}$.
Trivially, there are polynomials ${\cal N}(\tau_1,\dots,\tau_M)$ of
degree $d$ with a number of rational zeros at least equal to the number of
coefficients minus $1$, that is, equal to 
$\bar N_P\equiv M!^{-1}\prod_{k=1}^M(d+k)-1\sim d^M/M!+O(d^{M-1})$.
For $M=1$, this corresponds to take the univariate polynomial 
\be\label{univ_poly}
{\cal N}(\tau)=(\tau-x_1)(\tau-x_2)\dots(\tau-x_d).
\ee
A better construction of a multivariate polynomial is a generalization 
of the univariate polynomial in Eq.~(\ref{univ_poly}).
Given linear functions $L_{i,s}(\vec\tau)$, the polynomial 
$$
\tilde P=\sum_{i=1}^M\prod_{s=1}^d L_{i,s}(\vec\tau)
$$
has generally a number of rational points $N_P$ at least equal to $d^M$, which
is the square root of the upper bound, up to a constant.
For $d<4$ and $M=2$, the number of rational zeros turns 
out to be infinite, since the genus is smaller than $2$ 
(see Sec.~\ref{sec_infinite_points} for the case of
infinite rational points).
A naive computation of $\tilde P(\vec\tau)$ takes $d M^2$ 
arithmetic operations, that is, its complexity is polynomial in $M$.
This example provides an illustration of the complexity test described 
previously in Claim~\ref{claim1}. Expressing
$d$ in terms of $M$ and $\xi=\log N_P$ and assuming that 
${\bf C}_0\sim d M^2$, we have that
$$
{\bf C}_0(\xi)=M^2 e^{M^{-1}\xi},
$$
which is subexponential in $\xi$ (provided that $M$ is a growing function of $\xi$), 
which 
is a necessary condition for a subexponential algorithm. However, the polynomial 
does not pass the litmus test, as ${\bf C}_0(\xi)^M$ grows exponentially.
\subsubsection{The case of infinite rational zeros}
\label{sec_infinite_points}
Until now, we have assumed that the rational function has a finite number of rational
zeros over the rationals. However, in the multivariate case, it is possible to 
have non-zero functions
with an infinite number of zeros. For example, this is the case of a bivariate
polynomials with genus equal to zero and one, which correspond to parametrizable
curves and elliptic curves, respectively. We can also have functions whose
zero locus contains linear subspaces with positive dimension, which can have 
infinite rational points. Since the
probability of having ${\cal R}$ equal to zero modulo $p$ increases with the number
of zeros over the rationals, this would imply that the probability is equal
to $1$ if the number of zeros is infinite. This is not evidently the case. For
example, if ${\cal R}$ is zero for $x_1=0$ and $M>1$, evidently the function
has infinite rational points over $\mathbb{Q}$, but the number of points with
$x_1=0$ over $\mathbb{Z}_p$ is $p^{M-1}$, which is $p$ times less than the
number of points in the space.
Once again,
we are interested more properly to sets of $N_P$ rational zeros over $\mathbb{Q}$
such that a large fraction of them remain distinct over finite fields whose order 
is greater than about $N_P^{1/M}$. Under this condition, $N_P$ cannot be infinite
and is constrained by Ineq.~(\ref{up_bound}). If there are linear subspaces with
dimension $h>0$ in the zero locus of $\cal R$, we may fix some of the parameters 
$\vec\tau$, so that
these spaces become points. In the next sections, we will build rational functions
having isolated rational points and possible linear subspaces in the zero locus. 
If there are subspaces with dimension $k>0$ giving a dominant contribution to
factorization, we can transform them to isolated rational
points by fixing some parameters without changing the asymptotic complexity of
the algorithm.
Isolated rational points are the only relevant points for an asymptotic study of 
the complexity of the factoring algorithm, up to a dimension reduction. Thus,
we will consider only them and will not care of the other linear subspaces.

\section{Setting the problem in the framework of algebraic geometry}
\label{sec_arg_geo}

Since the number of zeros $N_P$ is constrained by Ineq.~(\ref{up_bound}),
the complexity of computing the rational function ${\cal R}(\vec\tau)$ must be 
subexponential or polynomial in $\log d$ in order to have ${\bf C}_0(\xi)$ 
subexponential or polynomial. This complexity scaling is attained if, for
example, $\cal R$ is a polynomial with few monomials.
The univariate polynomial $P=\tau^d-1$, which is pivotal in Pollard's $p-1$
algorithm, can be evaluated with a number of arithmetic operations scaling 
polynomially in $\log d$. This is achieved by consecutively applying
polynomial maps. For example, if $d=2^g$, then $\tau^d$ is computed through
$g$ applications of the map $x\rightarrow x^2$ by starting with $x=\tau$.
However, polynomials with few terms have generally few zeros over 
$\mathbb{Q}$. More general polynomials and rational functions with friendly 
computational complexity are obviously available and are obtained by consecutive
applications of simple functions, as done for $\tau^d-1$. This leads us to formulate
the factorization problem in the framework of algebraic geometry as
an intersection problem. 

\subsection{Intersection points between a parametrizable variety and 
a hypersurface}
\label{sec_intersection}

Considering only the operations defined in the field, the most general rational functions
${\cal R}(\vec\tau)$ with low complexity can be evaluated through the consecutive
application of a small set of simple rational equations of the form
\be\label{ratio_eqs}
\begin{array}{c}
x_{n-M}=\frac{{\cal N}_{n-M}(x_{n-M+1},\dots,x_n)}{{\cal D}_{n-M}(x_{n-M+1},\dots,x_n)} \\
x_{n-M-1}=\frac{{\cal N}_{n-M-1}(x_{n-M},\dots,x_n)}{{\cal D}_{n-M-1}(x_{n-M},\dots,x_n)} \\
\vdots \\
x_1=\frac{{\cal N}_1(x_2,\dots,x_n)}{{\cal D}_1(x_2,\dots,x_n)} \\
{\cal R}=P_0(x_1,\dots,x_n),
\end{array}
\ee
where $P_0$ is a polynomial. If the numerators and denominators ${\cal N}_k$ and ${\cal D}_k$
do not contain too many monomials, then the computation of ${\cal N}_k/{\cal D}_k$ can
be performed efficiently. Assuming that the computational complexity
of these rational functions is polynomial in $n$, the complexity of $\cal R$ is 
polynomial in $n$. The computation of ${\cal R}(\vec\tau)$ is performed by setting the
last $M$ components $x_{n-M+1},\dots,x_n$ equal to $\tau_1,\dots,\tau_M$
and generating the sequence $x_{n-M},x_{n-M-1},\dots,x_1,{\cal R}$ according
to Eqs.~(\ref{ratio_eqs}), which ends up with the value of ${\cal R}$.
The procedure may fail to compute the
right value of ${\cal R}(\vec\tau)$ if some denominator 
${\cal D}_k(\vec\tau)\equiv {\cal D}_k[x_{k+1}(\vec\tau),\dots,x_n(\vec\tau)]$
turns out to be equal to zero during the computation. However, since
our only purpose is to generate the zero of the field, we can take a zero divisor 
as outcome and stop the computation of the sequence. In this way, the
algorithm generates a modified function ${\cal R}'(\vec\tau)$.
Defining $\bar{\cal N}_k(\vec\tau)$ as the numerator of the rational function
${\cal D}_k(\vec\tau)$, we have
\be 
{\cal R}'(\vec\tau)=\left\{
\begin{array}{lr}
{\cal R}(\vec\tau)  & \;\; \text{if}\;\;  \bar{\cal N}_1(\vec\tau)\dots \bar{\cal N}_{n-M}(\vec\tau)\ne0 \\
0  & \text{otherwise}
\end{array}
\right.
\ee
The function ${\cal R}'(\vec\tau)$ has the zeros of
${\cal N}(\vec\tau)\bar{\cal N}_1(\vec\tau)\dots \bar{\cal N}_{n-M}(\vec\tau)$.
For later reference, let us define the following.
\begin{algorithm} 
\label{algo1}
Computation of ${\cal R}'(\vec\tau)$.
\begin{enumerate}
\item set $(x_{n-M+1},\dots,x_n)=(\tau_1,\dots,\tau_M)$; 
\item set $k=n-M>0$;
\item\label{attr} set
$x_k=\frac{{\cal N}_k(x_{k+1},\dots,x_n)}{{\cal D}_k(x_{k+1},\dots,x_n)}$.
If the division fails, return the denominator  as outcome;
\item set $k=k-1$;
\item\label{last_step}
if $k=0$, return $P_0(x_1,\dots,x_n)$, otherwise go back to \ref{attr}.
\end{enumerate}
\end{algorithm}
The zeros of the denominators are not expected to give an effective
contribution on the asymptotic complexity of the factoring algorithm, otherwise it
would be more convenient to reduce the number of steps of the sequence by one and
replace the last function $P_0$ with the denominator ${\cal D}_1$. Let us show that.
Let us denote by $N_1$ the number of rational zeros of ${\cal R}'$ over some
finite field with all the denominators different from zeros. They are the zeros
returned at step~\ref{last_step} of Algorithm~\ref{algo1}.
Let $N_T$ be the total number of zeros. We remind that the factoring complexity
is about $p^{k M}$ times the ratio between the complexity of $\cal R$ and the number
of zeros. If the algorithm is more
effective than the one with one step less and $P_0$ replaced with
${\cal D}_1$, then $\frac{C_T}{N_T}<\frac{C_T-C_1}{N_T-N_1}$. 
where $C_1$ and $C_T$ are the number of arithmetic operations of the last
step and of the whole algorithm, respectively. Since $C_1\ge1$, we have
$$
N_1>N_T C_T^{-1}
$$
In order to have a subexponential
factoring algorithm, $C_T$ must scale subexponentially 
in $\log N_T$. Thus,
$$
N_1>N_T e^{-\alpha (\log N_T)^\beta}
$$
for some positive $\alpha$ and $0<\beta<1$. That is,
$$
\log N_1>\log N_T -\alpha (\log N_T)^\beta.
$$
If we assume polynomial complexity, we get the tighter bound
$$
\log N_1>\log N_T -\alpha \log\log N_T.
$$
These inequalities imply that the asymptotic complexity of the factoring
algorithm does not change if we discard the zero divisors at Step~\ref{attr}
in Algorithm~\ref{algo1}. Thus, for a complexity analysis, we can consider only
the zeros of ${\cal R}'(\vec\tau)$ with all the denominators ${\cal D}_k(\vec\tau)$
different from zero. This will simplify the subsequent discussion. 
Each of these zeros are associated with an $n$-tuple $(x_1,\dots,x_n)$,
generated by Algorithm~\ref{algo1} and solutions of Eqs.~(\ref{ratio_eqs}).
Let us denote by ${\cal Z}_P$ the set of these $n$-tuples.

By definition, an element in ${\cal Z}_P$ is a zero of the set of $M-n+1$ polynomials
\be
\label{poly_affine}
\begin{array}{l}
P_0(x_1,\dots,x_n),  \\
P_k(x_1,\dots,x_n)=x_k {\cal D}_k(x_{k+1},\dots,x_n)-{\cal N}_k(x_{k+1},\dots,x_n), \;\;\;
k\in\{1,\dots,n-M\}.
\end{array}
\ee
The last $n-M$ polynomials define an algebraic set of points, say $\cal A$,
having one irreducible branch parametrizable by Eqs.~(\ref{ratio_eqs}).
This branch defines an algebraic variety which we denote by $\cal V$. 
The algebraic set may have other irreducible components which do not care about.
The polynomial $P_0$ defines a hypersurface, say $\cal H$. Thus, the
set ${\cal Z}_P$ is contained in the intersection between $\cal V$ and $\cal H$. This 
intersection may contain singular points of $\cal V$ with ${\cal D}_k(x_{k+1},\dots,x_n)=0$
for some $k$, which are not relevant for a complexity analysis, as shown previously.
Thus, the factorization problem  is reduced to search for non-singular rational points
of a parametrizable variety $\cal V$, whose intersection with a hypersurface
$\cal H$ contains an arbitrarily large number $N_P$ of rational points.
If $N_P$ and ${\bf C}_0$ scale exponentially and polynomially 
in the space dimension $n$, respectively, then the complexity of factorization is polynomial,
provided that the number of parameters $M$ scales sublinearly as a power of $n$.
In the limit case of 
\be\label{subexp_cond}
M\sim n/(\log n)^\beta
\ee
with $\beta>1$,
the complexity scales subexponentially as $e^{b (\log p)^{1/\beta}}$. Thus, if
$M$ has the scaling property~(\ref{subexp_cond}) with $\beta>3$, then there
is an algorithm outperforming asymptotically the general number field sieve.
A subexponential computational complexity is also obtained if the complexity
of evaluating a point in $\cal V$ is subexponential. 

The parametrization of $\cal V$ is a particular case of rational parametrization
of a variety. We call it \emph{Gaussian parametrization} since the triangular form of 
the polynomials $P_1,\dots,P_{n-M}$ resembles Gaussian elimination.
Note that this form is invariant under the transformation
\be\label{poly_replacement}
P_k\rightarrow P_k+\sum_{k'=k+1}^{n-M}\omega_{k,k'} P_{k'}.
\ee
The form is also invariant under the variable
transformation
\be\label{invar_trans}
x_k\rightarrow x_k+\sum_{k'=k+1}^{n+1} \eta_{k,k'} x_{k'}
\ee
with $x_{n+1}=1$.

It is interesting to note that if $N_P'$ out of the $N_P$ points in ${\cal Z}_P$
are collinear, then it is possible to build another variety with Gaussian
parametrization and a hypersurface over a $(n-1)$-dimensional subspace such that
their intersection contains the $N_P'$ points. For later reference, let us
state the following.
\begin{lemma}
\label{lemma_dim_red}
Let ${\cal Z}_P$ be the set of common zeros of the polynomials~(\ref{poly_affine})
with ${\cal D}_k(x_{k+1},\dots,x_n)\ne 0$ over ${\cal Z}_P$ for $k\in\{1,\dots,n-M\}$.
If $N_P'$ points in ${\cal Z}_P$ are solutions of the linear equation $L(x_1,\dots,x_n)=0$,
then there is a variety with Gaussian parametrization and a hypersurface over
the  $(n-1)$-dimensional subspace defined by $L(x_1,\dots,x_n)=0$ such that their 
intersection contains the $N_P'$ points.
\end{lemma}
{\it Proof.} 
Given the linear function $L(x_1,\dots,x_n)\equiv l_{n+1}+\sum_{k=1}^n l_k x_k$,
let us first consider the case with $l_k=0$ for $k\in\{1,\dots,n-M\}$. Using
the constraint $L=0$, we can set one of the $M$ variables $x_{n-M+1},\dots,x_n$ 
as a linear function of the remaining $M-1$ variables. Thus, we get a new
set of polynomials retaining the original triangular form. The new parametrizable
variety, say ${\cal V}'$, has $M-1$ parameters. The intersection of ${\cal V}'$ with
the new hypersurface over the $(n-1)$-dimensional space contains the $N_P'$ points.
Let us now consider the case with $l_k=0$ for $k\in\{1,\dots,\bar k\}$, where
$\bar k$ is some integer between $0$ and $n-M-1$, such that $l_{\bar k+1}\ne 0$.
We can use the constraint $L=0$ to set $x_{\bar k+1}$ as a linear function
of $x_{\bar k+2},\dots,x_n$. 
We discard the polynomial 
$P_{\bar k+1}$ and eliminate the $(\bar k+1)$-th variable from the remaining polynomials.
We get $n-1$ polynomials retaining the original triangular form in $n-1$ variables
$x_1,\dots,x_{\bar k},x_{{\bar k}+2},\dots,x_n$. The intersection between the
new parametrizable variety and the new hypersurface contain the $N_P'$ points
$\square$.
\newline
This simple lemma will turn out to be a useful tool in different parts of the
paper.

In Section~\ref{sec_common_zeros}, we show how to build a set of polynomials $P_k$
with a given number $N_P$ of common rational zeros by using some tools of algebraic geometry 
described in Appendix~\ref{alg_geom}. 
In Sec.~\ref{build_up}, we close the circle by 
imposing the form~(\ref{poly_affine}) for the polynomials $P_k$ with
the constraint that ${\cal D}_k(x_{k+1},\dots,x_n)\ne 0$ for $k\in\{1,\dots,n-M\}$ over
the set of $N_P$ points.

\subsection{Sets of polynomials with a given number of zeros over a number field}
\label{sec_common_zeros}

In this subsection, we build polynomials with given $N_P$ common rational zeros 
as elements of an ideal $I$ generated by products of linear functions. This construction 
is the most general.
The relevant information on the ideal $I$ is summarized by a satisfiability 
formula in conjunctive normal form without negations and a linear matroid.
The formula and the matroid uniquely determine the number $N_P$ of rational common zeros
of the ideal. Incidentally, we also show that the information can be encoded in a more 
general formula with negations by a suitable choice of the matroid.

Every finite set of points in an $n$-dimensional space is an algebraic set, that 
is, they are all the zeros of some set of polynomials. More generally, the union
of every finite set of linear subspaces is an algebraic set. 
In the following, we 
will denote linear polynomials by a symbol with a hat; namely, $\hat a$ is meant 
as $a_{n+1}+\sum_{i=1}^n a_i x_i$. Let us denote by $\vec x$ the $(n+1)$-dimensional
vector $(x_1,\dots,x_n,x_{n+1})$, where $x_{n+1}$ is an extra-component that is 
set equal to $1$. A linear polynomial $\hat a$ is written in the form
$\vec  a\cdot\vec x$.
Let $V_1,\dots, V_{L}$ be a set of linear subspaces and $I_1,\dots,I_L$ their 
associated radical ideals. The codimension of the $k$-th subspace is denoted by $n_k$. 
The minimal set of generators of the $k$-th ideal contains $n_k$ independent
linear polynomials, say $\hat a_{k,1},\dots,\hat a_{k,n_k}$, so that
\be
\vec x\in V_k \Leftrightarrow\vec a_{k,i}\cdot\vec x=0\;\; \forall i\in\{1,\dots,n_k\}.
\ee
If the codimension $n_k$ is equal to $n$, then $V_k$ contains one
point. We are mainly interested to these points, whose number is
taken equal to $N_P$. The contribution of higher dimensional subspaces to 
the asymptotic complexity of the factoring algorithm is irrelevant up to a dimension 
reduction (see also Sec.~\ref{sec_infinite_points} and the remark in the
end of the section). Since only isolated points are relevant, we could just consider
ideals whose zero loci contain only isolated points. However,
we allow for the possible presence of subspaces with positive dimension
since they may simplify the set of the generators or the form of the polynomials
$P_k$ that eventually we want to build.

Let $\cal Z$ be the union of the subspaces $V_k$. The product 
$I_1\cdot I_2\cdot\dots I_L\equiv \tilde I$ is associated with $\cal Z$, that is,
${\cal Z}={\bf V}(\tilde I)$. A set of generators of the ideal $\tilde I$ is 
\be
\prod_{k=1}^{L}\hat a_{k,i_k}\equiv G_{i_1,\dots,i_L}(\vec x) \;\;\;
i_r\in\{1,\dots,n_r\}, r\in\{1,\dots,L\}. 
\ee 
Thus, we have that
\be
\vec x\in {\cal Z} \Leftrightarrow G_{i_1,\dots,i_L}(\vec x)=0
\;\;\;\; i_r\in\{1,\dots,n_r\}, r\in\{1,\dots,L\}.
\ee
Polynomials in the ideal $\tilde I$ are zero in the set $\cal Z$.
This construction is not the most general, as $\tilde I$  is not radical.
Thus, there are polynomials that are not in $\tilde I$, but 
their zero locus contains $\cal Z$.
Furthermore, the number of generators and the number of their 
factors grow polynomially and linearly in $N_P$, respectively. 
This makes it hard to build polynomials in $\tilde I$ whose
complexity is  polynomial in $\log N_P$.
The radicalization of the ideal and the assumption of
special arrangements of the subspaces in $\cal Z$ 
can reduce drastically both the degree of the generators and
their number. For example, let us assume that $V_1$ and $V_2$
are two isolated points in the $n$-dimensional space and, thus, 
$n_1=n_2=n$. The overall number of generators $a_{1,1},\dots,a_{1,n}$ and 
$a_{2,1},\dots,a_{2,n}$ is equal to $2 n$. Thus, there are
$n-1$ linear constraints among the generators. Using linear
transformations, we can write these constraints as
$$
\hat a_{1,i}=\hat a_{2,i}\equiv\hat a_i \;\;\;\forall i\in\{2,\dots,n\}.
$$
Every generator $G_{i,i,i_3,\dots,i_L}$ with $i\ne 1$ is equal
to $\bar G_{i,i_3,\dots,i_L}=a_i^2 \prod_{k=3}^L \hat a_{k,i_k}$. The polynomial
$\bar G'_{i,i_3,\dots,i_L}\equiv a_i \prod_{k=3}^L \hat a_{k,i_k}$ is not an element of the
ideal $\tilde I$, but it is an element of its radical. Indeed,
it is zero in the algebraic set $\cal Z$. Thus, we extend
the ideal by adding these new elements. This extension allows
us to eliminate all the generators $G_{i_1,i_2,\dots,i_L}$ with
$i_1=1$ or $i_2=1$, since they are generated by $\bar G'_{i,i_3,\dots,i_L}$.
Thus, the new ideal has the generators,
\begin{equation}
\left.
\begin{array}{l}
\hat a_{1,1}\hat a_{2,1}\prod_{k=3}^L a_{k,i_k}  \\
\hat a_i\prod_{k=3}^L a_{k,i_k},\;\;\;  i\in\{2,\dots,n\}
\end{array}\right\}  
i_r\in\{1,\dots,n_r\}, r\in\{3,\dots,L\}. 
\end{equation}
Initially, we had $n^2\prod_{k=3}^{L}n_k$ generators. Now, their
number is $n\prod_{k=3}^{L}n_k$. A large fraction of them has the
degree reduced by one. We can proceed with the other points and
further reduce 
both degrees and number of generators. Evidently,
this procedure cannot take to a drastic simplification of the
generators if the points in $\cal Z$ are in general position, since
the generators must contain the information about these positions.
A simplification is possible if the points have special arrangements
taking to contraction of a large number of factors in the generators.
Namely, coplanarity of points is the key feature that can take to a
drastic simplification of the generators.
In a $n$-dimensional space, there are at most $n$ coplanar points
in general position. Let us consider algebraic sets containing
larger groups of coplanar points. For example, let us assume that 
the first $m$ sets $V_1,\dots,V_m$ are distinct coplanar points, with 
$m\gg n$. Then, there is a vector $\vec a_1$ such that $\vec a_1\cdot\vec x=0$ 
for every $\vec x$ in the union of the first $m$ linear spaces. It is convenient
to choose the linear polynomial $\vec a_1\cdot\vec x$ as common generator of 
the first $m$ ideals $I_1,\dots,I_m$. Let us set $\hat a_{k,1}=\hat a_1$
for $k\in\{1,\dots,m\}$. Every generator $G_{i_1,\dots,i_L}$ with $i_k=1$
for some $k\in\{1,\dots,m\}$ is contracted to a generator of the form
$\hat a\prod_{k=m+1}^L \hat a_{k,i_k}$. If there are other groups of
coplanar points, we can perform other contractions. 
\begin{definition}
Given an integer ${\bar n}>n$, we define $\Gamma_s$ with $s\in\{1,\dots,{\bar n}\}$ as
a set of $s$-tuples $(i_1,\dots,i_s)\in\{1,\dots,{\bar n}\}^s$ 
with $i_k < i_{k'}$ for $k < k'$. That is, 
\begin{equation}
\forall s\in\{1,\dots,{\bar n}\},\;\;\; 
\Gamma_s\subseteq \{(i_1,\dots,i_s)\in\{1,\dots,{\bar n}\}^s|  i_k < i_{k'},\forall k, k' \text{ s.t. }
k<k' \}.
\end{equation}
\end{definition}
The final result of the inclusion of elements of the radical ideal is another 
ideal, say $I$, with generators of the form
\be\label{generators}
\begin{array}{l}
\hat a_{i_1}  \;\;\;\forall i_1\in\Gamma_1  \\
\hat a_{i_1}\hat a_{i_2}  \;\;\;\forall (i_1,i_2)\in\Gamma_2  \\
\dots   \\
\hat a_{i_1}\hat a_{i_2}\dots\hat a_{i_{{\bar n}}}  \;\;\;\forall (i_1,i_2,\dots,i_{{\bar n}})\in\Gamma_{{\bar n}},
\end{array}
\ee
where $\{\hat a_1,\dots,\hat a_{{\bar n}}\}\equiv\Phi$ is a set of ${\bar n}$ linear polynomials.
Polynomials in this form generate the most general ideals whose zero loci contain a given finite set
of points. This is formalized by the following.
\begin{lemma}
Every radical ideal associated with a finite set ${\cal Z}_P$ of points is generated by a set of 
polynomials of form~(\ref{generators}) for some $\bar n$. 
\end{lemma}
{\it Proof}. This can be shown with a naive construction. Given a set $\bar S$ of $N_P$ points 
associated with the ideals $I_1,\dots,I_{N_P}$, the product
$I_1\cdot \dots I_{N_P}$ is an ideal associated with the set $\bar S$, which can be radicalized by
adding a certain number of univariate square-free polynomials as generators~\cite{seidenberg}. 
The resulting ideal is generated by a set of polynomials of form~(\ref{generators}). $\square$
\newline
With the construction used in the proof, $\bar n$ ends up to be equal to the number of points 
in ${\cal Z}_P$, which is not optimal for our purposes. We are interested to keep $\bar n$ 
sufficiently small, possibly scaling polynomially in the dimension $n$. 
This is possible only if the points in the zero locus
have a `high degree' of collinearity.
Thus, a bound on $\bar n$ sets a restriction on ${\cal Z}$.

The minimal information on $\Phi$ that is relevant for determining the number $N_P$ of points in
$\cal Z$ is encoded in a linear \emph{matroid}, of which $\Phi$ is one linear representation.
Thus, the sets $\Gamma_s$ and the matroid determine $N_P$.
Note that the last set $\Gamma_{{\bar n}}$ has at most one element. The linear generators
can be eliminated by reducing the dimension of the affine space, see Lemma~\ref{lemma_dim_red}.
Thus, we can set $\Gamma_1=\emptyset$. Every subset $\Phi_{sub}$ of $\Phi$ is associated with a
linear space $V_{sub}$ whose points are the common zeros of the linear functions in $\Phi_{sub}$.
That is, $V_{sub}={\bf V}(I_{sub})$, where $I_{sub}$ is the ideal generated by $\Phi_{sub}$.
Let us denote briefly by ${\bf V}(\Phi_{sub})$ the linear space ${\bf V}(I_{sub})$.
This mapping from subsets of $\Phi$ to subspaces is not generally injective. Let 
$\hat a'\in\Phi\setminus \Phi_{sub}$ be a linear superposition of the functions in 
$\Phi_{sub}$, then $\Phi_{sub}$ and $\Phi_{sub}\cup\{\hat a'\}$ represent the same linear space. 
An injective mapping is obtained by considering only the maximal subset associated with a linear 
subspace. These maximal subsets are called \emph{flats} in matroid theory.
\begin{definition}
\emph{Flats} of the linear matroid $\Phi$ are defined as subsets
$\Phi_{sub}\subseteq \Phi$ such that 
no function in $\Phi\setminus \Phi_{sub}$ is
linearly dependent on the functions in $\Phi_{sub}$.
\end{definition}
Let us also define the closure of a subset of $\Phi$.
\begin{definition}
Given a subset $\Phi_{sub}$ of $\Phi$ associated with subspace $V$, its closure
$\text{cl}(\Phi_{sub})$ is the flat associated with $V$.
\end{definition}
The number of independent functions in a flat is called \emph{rank} of the flat.
The whole set $\Phi$ is a flat of rank $n+1$, which is associated with an empty space. Flats 
of rank $n$ define points of the $n$-dimensional affine space (with $x_{n+1}=1$).
More generally,
flats of rank $k$ define linear spaces of dimension $n-k$.
The dimension of a flat $\Phi_{flat}$ is meant as the dimension of the space
${\bf V}(\Phi_{flat})$.

The structure of the generators~(\ref{generators}) resembles 
a Boolean satisfiability problem (SAT) in conjunctive normal form without negations. 
Let us interpret $\hat a_i$ as a logical variable $a_i$ which is $\true$ or $\false$ if
the function is zero or different from zero, respectively. Every subset 
$\Phi_{sub}\subseteq\Phi$ is identified with a string $(a_1,\dots,a_{{\bar n}})$ such
that $a_i=\true$ if and only if $\hat a_i\in\Phi_{sub}$.

The SAT formula associated with the generators~(\ref{generators}) is
\be\label{SAT}
\bigwedge\limits_{k=2}^{{\bar n}} \left( \bigvee\limits_{i\in  \Gamma_k  } a_i \right).
\ee
Given a flat $\Phi_{flat}$, the linear space ${\bf V}(\Phi_{flat})$
is a subset of $\cal Z$ if and only if $\Phi_{flat}$
is solution of the SAT formula. If a set $\Phi_{sub}\subseteq\Phi$ is solution
of the SAT formula, then the flat $\text{cl}(\Phi_{sub})$ is also solution
of the formula. Thus, satisfiability implies that there are flats as solutions
of the formula. This does not mean that
satisfiability implies that
$\cal Z$ is non-empty. Indeed, if the dimension of ${\bf V}(\Phi_{sub})$ is
negative for every solution $\Phi_{sub}$, then the set $\Phi$ is the only
flat solution of the formula.
We are interested to the isolated points in $\cal Z$. 
A point $p\in{\cal Z}$
is \emph{isolated} if there is a SAT solution $\Phi_{flat}$ with zero dimension 
such that $p\in{\bf V}(\Phi_{flat})$ and no flat $\Phi_{flat}'\subset\Phi_{flat}$ is
solution of the Boolean formula. We denote by ${\cal Z}_P$ the subset in $\cal Z$
containing the isolated points. Since the number $N_P$ of isolated points is completely
determined by the SAT formula and the linear matroid, the information on these
two latter objects is the most relevant. Given them, the linear functions $\hat a_i$
have some free coefficients.

{\bf Remark}. In general, we do not rule out sets $\cal Z$ 
containing subspaces with positive dimension, however these subspaces are irrelevant
for the complexity analysis of the factoring algorithm. For example, if subspaces of 
dimension $d_s<M$ 
give a dominant contribution to factorization, then we can generally eliminate $d_s$ out of 
the $M$ parameters by setting them equal to constants,
so that the subspaces are reduced to points. Furthermore, 
subspaces with dimension greater than $M-1$ are not in the parametrizable variety $\cal V$,
whose dimension is $M$. Neither the overall contribution of all the subspaces with positive
dimension can provide a significant change in the asymptotic complexity up to
parameter deletions. Thus, only isolated points of 
$\cal Z$ are counted without loss of generality.

\subsection{Boolean satisfiability and algebraic-set membership}
As we said previously,
the Boolean formula does not encode all the information about the number of isolated
points in ${\cal Z}$, which also depends on the independence relations among the vectors 
$\vec a_i$, specified by the matroid. A better link between the SAT problem
and the membership to ${\cal Z}$ can be obtained if we consider sets $\Phi$ 
with cardinality equal to $2 n$ and interpret half of the
functions in $\Phi$ as negations of the others. Let us
denote by $\hat a_{0,1},\dots,\hat a_{0,n}$ and $\hat a_{1,1},\dots,\hat a_{1,n}$ 
the $2n$ linear functions of $\Phi$. For general functions, we have the following.
\begin{property}
\label{indep_functions}
The set of vectors $\{\vec a_{s_1,1},\dots,\vec a_{s_n,n},\vec a_{1-s_k,k}\}$
is independent for every string $\vec s=(s_1,\dots,s_n)\in\{0,1\}^n$ and every 
$k\in\{1,\dots,n\}$.
\end{property}
This property generally holds if the functions are picked up at random. 
Let us assume that $\Phi$ satisfies Property~\ref{indep_functions}. This 
implies that  $\{\hat a_{s_1,1},\dots,\hat a_{s_n,n}\}$ are linearly independent
and equal to zero at one point $\vec x_{\vec s}$. Furthermore,
Property~\ref{indep_functions} also implies that different strings $\vec s$ are
associated with different points $\vec x_{\vec s}$.
\begin{lemma}
\label{lemma_indep}
Let $\{\vec a_{0,1},\dots,\vec a_{0,n},\vec a_{1,1},\dots,\vec a_{1,n}\}$ be a set
of $2n$ vectors  satisfying Property~\ref{indep_functions}. Let $\vec x_{\vec s}$
be the solution of the equations $\hat a_{s_1,1}=\dots=\hat a_{s_n,n}=0$. 
If $\vec s\ne\vec r$, then $\vec x_{\vec s}\ne\vec x_{\vec r}$.
\end{lemma}
{\it Proof}. Let us assume that $\vec x_{\vec s}=\vec x_{\vec r}$ with $\vec s\ne\vec r$. 
There is
an integer $k\in\{1,\dots,n\}$ such that $s_k\ne r_k$. Thus, the set of
vectors $\{\vec a_{s_1,1},\dots,\vec a_{s_n,n},\vec a_{1-s_k,k}\}$ 
are orthogonal to $\vec x_{\vec s}$. Since the dimension of the vector space
is $n+1$, the set of $n+1$ vectors are linearly dependent, in contradiction
with the hypotheses. $\square$

Now, let us define the set ${\cal Z}$ as the zero locus of the ideal generators
\be\label{generators2}
\begin{array}{l}
\hat a_{0,i} \hat a_{1,i}  \;\;\;\forall i\in\{1,\dots,n\}  \\
\hat a_{s_1,i_1}\hat a_{s_2,i_2}  \;\;\;\forall (s_1,i_1;s_2,i_2)\in\Gamma_2  \\
\dots   \\
\hat a_{s_1,i_1}\hat a_{s_2,i_2}\dots\hat a_{s_n,i_n}  \;\;\;\forall (s_1,i_1;s_2,i_2,\dots,s_n,i_n)\in\Gamma_n.
\end{array}
\ee
The first $n$ generators provide an interpretation of $\hat a_{1,i}$ as negation
of $\hat a_{0,i}$, as consequence of Property~\ref{indep_functions}. The $i$-th generator
implies that $(a_{0,i},a_{1,i})$ is equal to $(\true,\false)$, $(\false,\true)$
or $(\true,\true)$. However, the last case is forbidden by Property~\ref{indep_functions}.
Assume that  $(a_{0,i},a_{1,i})$ is equal to $(\true,\true)$ for some $i$.
Then, there would be $n+1$ functions $\hat a_{s_1,1},\dots,\hat a_{s_n,n},\hat a_{1-s_i,i}$
equal to zero, which is impossible since they are independent. Thus, the algebraic
set defined by the first $n$ generators contains $2^n$ distinct points, as implied
by Lemma~\ref{lemma_indep}, which are associated with all
the possible states taken by the logical variables. The remaining
generators set further constraints on these variables and define a Boolean formula
in conjunctive normal form. With this construction there is a one-to-one correspondence
between the points of the algebraic set ${\cal Z}$ and the solutions of the Boolean
formula.

There is a generalization of the generators~(\ref{generators2}) that allows us
to weaken Property~\ref{indep_functions} while retaining the one-to-one correspondence.
Let $R_1,\dots,R_m$ be $m$ disjoint non-empty sets such that
$\cup_{k=1}^m R_k=\{1,\dots,n\}$. 
\begin{property}
\label{indep_functions2}
The set of vectors  $\cup_{k=1}^m\{\vec a_{s_k,i}|i\in R_k\}\equiv A_{\vec s}$ is 
independent for every $\vec s=(s_1,\dots,s_m)\in\{0,1\}^m$. Furthermore,
every vector $\vec a_{s,i}\notin A_{\vec s}$ is not in $\text{span}(A_{\vec s})$,
with $s\in\{0,1\}$ and $i\in\{1,\dots,n\}$.
\end{property}
\begin{lemma}
\label{lemma_indep2}
Let $\{\vec a_{0,1},\dots,\vec a_{0,n},\vec a_{1,1},\dots,\vec a_{1,n}\}$ be a set
of $2n$ vectors  satisfying Property~\ref{indep_functions2}. Let $\vec x_{\vec s}$
be the solution of the equations 
$$
\hat a_{s_k,i}=0 \;\;\; i\in R_k, k\in\{1,\dots,m\}
$$
for every  $\vec s\in\{0,1\}^m$.
If $\vec s\ne\vec r$, then $\vec x_{\vec s}\ne\vec x_{\vec r}$.
\end{lemma}
The generators~(\ref{generators2}) are generalized by replacing the first line 
with 
\be\label{block_gen}
\hat a_{0,i} \hat a_{1,j}, \;\;\;  (i,j)\in \cup_{k=1}^m (R_k \times  R_k).
\ee
Provided that Property~\ref{indep_functions2} holds there is a one-to-one correspondence
between the points in the algebraic set and the solutions of a SAT formula built according
to the following interpretation.
Each set of functions $\{\hat a_{0,i}|i\in R_k\}\equiv a_k$ is interpreted as a Boolean variable,
which is true if the functions in there are equal to zero. The set $\{\hat a_{1,i}|i\in R_k\}$ 
is interpreted as negation of $\{\hat a_{0,i}|i\in R_k\}$. The SAT formula is built in obvious
way from the set of generators. For example, the generator $\hat a_{0,i}\hat a_{0,j}$ with
$i\in R_{1}$ and $j\in R_{2}$ induces the clause $a_{1} a_{2}$. Different generators
can induce the same clause. Since the total number of solutions depends only on the
SAT formula, it is convenient to take the maximal set of generators compatible with
a given formula. That is, if $a_{1} a_{2}$ is a clause, then $\hat a_{0,i}\hat a_{0,j}$
is taken as a generator for every $i\in R_1$ and $j\in R_2$.

\subsection{$3$SAT-like generators}
\label{sec_3SAT}
SAT problems have clauses with an arbitrarily large number of literals. Special cases
are $2$SAT and $3$SAT, in which clauses have at most $2$ or $3$ literals. It is known
that every SAT problem can be converted to a $3$SAT one by increasing the number of
variables and replacing a clause with a certain number of smaller clauses containing
the new variables. For example, the clause $a\lor b\lor c\lor d$ can be replaced by
$a\lor b\lor x$ and $c\lor d\lor (\lnot x)$. An assignment satisfies the first clause
if and only if the other two clauses are satisfied for some $x$. An identical reduction 
can be performed also on the generators~(\ref{generators}). For example, a generator in $I$
of the form $\hat a_1\hat a_2\hat a_3\hat a_4\equiv G_0$ can be replaced by 
$\hat a_1\hat a_2 y\equiv G_1$, $\hat a_1\hat a_2(1-y)\equiv G_2$ 
and $y(1-y)\equiv G_3$, where $y$ is an additional variable. Also in this 
case, $G_0$ is equal to zero if and only if $G_1$  and $G_2$
are zero for $y=0,1$. Furthermore, the new extended ideal contains the
old one. Indeed, we have that $G_0=\hat a_3\hat a_4 G_1+\hat a_1\hat a_2 G_2$.

Note that all the polynomials in the ideal $I$ are independent of the
additional variables used in the reduction. Thus, if we build the 
polynomials~(\ref{poly_affine}) by using $3$SAT-like generators, then
all these polynomials may be independent of some variables.
Thus, we can consider generators in a $3$SAT form,
\be\label{generators_3SAT}
\begin{array}{l}
\hat a_{i_1}\hat a_{i_2}  \;\;\;\forall (i_1,i_2)\in\Gamma_2  \\
\hat a_{i_1}\hat a_{i_2}\hat a_{i_{3}}  \;\;\;\forall (i_1,i_2,i_3)\in\Gamma_{3}.
\end{array}
\ee
There is no loss of generality, provided that all the polynomials $P_k$ are 
possibly independent from $n_I$ variables. 

The number of isolated points satisfies the inequality
\be
\label{bound_N0_n}
N_P\le 3^n.
\ee
The actual number can be considerably smaller, depending on the
matroid and the number of clauses defining the Boolean formula.
The bound is attained if $n_c=3 n$, the generators have the
form $a_i b_i c_i$ with $i\in\{1,\dots,n\}$, and the independent 
sets of the matroid contain $n+1$ elements.
If there are only clauses with $2$ literals, then the bound is
\be
N_P\le 2^n,
\ee
which is strict if the generators have the form $a_i b_i$ with
$i\in\{1,\cdots,n\}$.
A consequence of these constraints is that the number $M$ of parameters
must scale sublinearly in $n$,
\be
\label{bound_M_n}
M\le K n^\beta, \;\;\;\;   0\le \beta<1
\ee
for some $K>0$.

\section{Building up the parametrizable variety and the hyperplane}
\label{build_up}
In this section, we put together the tools introduced previously to tackle our
problem of building the rational function ${\cal R}$ with the desired properties
of being computationally simple and having a sufficiently large set of zeros.
This problem has being reduced to the search of computationally simple 
polynomials $P_k$ of the form~(\ref{poly_affine}) with a number of 
common rational zeros growing sufficiently fast with the space dimension.
To build these polynomials, we first choose a set of generators of the
form~(\ref{generators}) such that the associated algebraic set $\cal Z$ has a set 
of $N_P$ points. Then, we write the polynomials $P_k$ as elements of the ideal
associated with $\cal Z$. Finally, we impose that the polynomials $P_k$ have the
form of Eqs.~(\ref{poly_affine}). 
\begin{procedure}
\label{procedure}
Building up of a parametrizable variety $\cal V$ with $M$ parameters
and $N_P$ intersection points.
\begin{enumerate}
\item Take a set of ${\bar n}$ unknown non-homogeneous linear functions in $n$ variables
with ${\bar n}>n$, say $\hat a_1,\dots,\hat a_{{\bar n}}$. Additionally, 
specify which set of vectors are linearly independent. In other words,
a linear matroid with ${\bar n}$ elements is defined. 
\item Choose and ideal $I$ with generators of the form~(\ref{generators_3SAT}) such that
the associated algebraic set $\cal Z$ contains a subset ${\cal Z}_P$ of
$N_P$ isolated points over some given number field.
\item Set the polynomials $P_s$ equal to elements of the ideal $I$
with $s\in\{0,\dots,n-M\}$. That is,
\begin{equation}
\label{poly_in_ideal}
P_s(\vec x)=\sum_{(i,j)\in \Gamma_2} C_{s,i,j}(\vec x) \hat a_i \hat a_j+
\sum_{(i,j,k)\in \Gamma_3} D_{s,i,j,k}(\vec x) \hat a_i \hat a_j\hat a_k,
\end{equation}
The polynomials $P_s$ with $s\in\{1,\dots,n-M\}$ define an algebraic set $\cal A$.
The polynomial $P_0$ defines a hyperplane $\cal H$. The number of parameters $M$ 
and the polynomial coefficients $C_{s,i,j}(\vec x)$ and $D_{s,i,j,k}(\vec x)$ are also unknown.
\item Search for values of the coefficients such that there is a parametrizable branch $\cal V$ 
in $\cal A$
with a number of parameters as small as possible. All the polynomials $P_s$ with 
$s\in\{0,\dots,n-M\}$  are possibly independent of some subset of variables (see Sec.~\ref{sec_3SAT}). 
The polynomials ${\cal D}_k$, as defined in Eq.~(\ref{poly_affine}) must be different
from zero in the set ${\cal Z}_P$.
\end{enumerate}
\end{procedure}
More explicitly, the last step leads us to the following.
\begin{problem}
\label{problem1}
Given the sets $\Gamma_2$ and $\Gamma_3$, and polynomials of the
form~(\ref{poly_in_ideal}), find linear functions 
$\hat a_1,\dots,\hat a_{{\bar n}}$ and coefficients $C_{s,i,j}(\vec x)$, 
$D_{s,i,j,k}(\vec x)$ such that
\be
\label{gauss_constrs}
\begin{array}{l}
\frac{\partial P_s}{\partial x_k}=0, \;\;\;\;  1\le k < s\le n-M, 
\vspace{2mm}
\\
\frac{\partial^2 P_s}{\partial x_s^2}=0, \;\;\;\; 1\le s\le n-M,
\vspace{2mm}
\\
\vec x\in{\cal Z}_P \Rightarrow \frac{\partial P_s}{\partial x_s}\equiv 
{\cal D}_s(x_{s+1},\dots,x_n)\ne 0,
\end{array}
\ee
under the constraint that $(\hat a_1,\dots,\hat a_{{\bar n}})$ is the representation
of a given matroid.
\end{problem}

{\bf Remark}.
If the algebraic set associated with the ideal $I$ is zero-dimensional,
this problem has always a solution for any $M$, since a rational univariate representation
always exists (see introduction). Essentially, the task is to find ideals such
that there is a solution with the coefficients $C_{s,i,j}(\vec x)$ and
$D_{s,i,j,k}(\vec x)$ as simple as possible, so that their computation is
efficient, given $\vec x$.

Let us remind that the constraints~(\ref{gauss_constrs}) are 
invariant under transformations~(\ref{poly_replacement},\ref{invar_trans}).
All the polynomials are possibly independent of a subset of $n_I$ 
variables, say $\{x_{n-n_I+1},\dots,x_{n}\}$,
\begin{equation}
\frac{\partial P_s}{\partial x_k}=0, \;\;\;\;  
\left\{
\begin{array}{l}
s\in\{0,\dots,n-M\} \\
k\in\{n-n_I+1,\dots,n\}
\end{array}\right.   
\end{equation}
These $n_I$ variables can be set equal to constants,
so that the actual number of significant parameters is $M-n_I$. 
The input of Problem~\ref{problem1} is given by a 3SAT formula
of form~(\ref{generators_3SAT}) and a linear matroid.
\begin{definition}
A 3SAT formula of form~(\ref{generators_3SAT}) and a linear matroid with
$\bar n$ elements is called a \emph{model}. 
\end{definition}
\noindent
In literature, the term `model' is occasionally used with a different 
meaning and refers to a solution of a SAT formula.

Problem~\ref{problem1} in its general form is quite intricate. First,
it requires the definition of a linear matroid and a SAT formula with an 
exponentially large number of solutions associated with isolated points.
Whereas it is easy to find examples of matroids and Boolean formulas
with this feature, it is not generally simple to characterize models
with an exponentially large number of isolated points.
Second, Eqs.~(\ref{gauss_constrs}) take to a large number of polynomial
equations in the unknown coefficients. Lemma~\ref{lemma_dim_red} can help
to reduce the search space by dimension reduction.
This will be shown in Sec.~\ref{sec_reduc_2SAT} with a simple example.
A good strategy is to start with simple models and low-degree coefficients
in Eq.~(\ref{poly_in_ideal}). In particular, we can take the coefficients
constant, as done later in Sec.~\ref{sec_quadr_poly}. This restriction
does not guarantees that Problem~\ref{problem1} has a solution for
a sufficiently small number of parameters $M$, but we can have some
hints on how to proceed.

\subsection{Required number of rational points vs space dimension}
Let assume that the computational complexity ${\bf C}_0$ of $\cal R$ 
is polynomial in the space dimension $n$, that is,
\begin{equation}
\label{ident_dim}
{\bf C}_0\sim n^{\alpha_0}.
\end{equation}
The factoring algorithm has polynomial complexity if
\be
\left.
\begin{array}{l}
K_1 n^\alpha\le \log N_P\le K_2 n \;\;\; 0<\alpha\le 1  \\
M\le (\log N_P)^\beta \;\;\; \beta<1
\end{array}
\right\}   \;\;\; (\text{polynomial complexity})
\ee
for $n$ sufficiently great, where $K_1$ is some positive
constant and $K_2=\log 3$. The upper bound is given 
by Eq.~(\ref{bound_N0_n}). The algorithm has subexponential
complexity 
${\bf C}\sim e^{b (\log N_P)^{\alpha}}$  with $0<\alpha<1$ if
\be
\left.
\begin{array}{r}
\log N_P\sim (\log n)^{1/\alpha}\;\;\; 0<\alpha<1,  \\
M\sim (\log n)^\frac{\beta}{\alpha}\;\;\; 0\le\beta<1-\alpha.
\end{array}
\right\} \;\;\;\;   (\text{subexponential complexity})
\ee
The upper bound on $\beta$ comes from Lemma~\ref{litmus}.
Thus, the number of rational points is required to
scale much less than exponentially for getting polynomial
or subexponential factoring complexity. Note that a slower
increase of $N_P$ induces stricter bounds on $M$ in terms
of $n$.

\subsection{Reduction of models}
\label{sec_reduc_2SAT}
In this subsection, we describe an example of model reduction. The model
reduction is based on Lemma~\ref{lemma_dim_red} and can be
useful for simplifying Problem~\ref{problem1}.
The task is to reduce a class of models  associated with an efficient
factoring algorithm to a class of simpler models taking to another
efficient algorithm, so that it is sufficient to search for solutions
of Problem~\ref{problem1} over the latter smaller class.

In our example, the matroid contains $2n$ elements and is
represented by the functions 
$\hat a_{0,1},\dots,\hat a_{0,n},\hat a_{1,1},\dots,\hat a_{1,n}$
satisfying Property~\ref{indep_functions}. 
\newline
{\bf Model A.}
\newline
Matroid with representation 
$(\hat a_{0,1},\dots,\hat a_{0,n},\hat a_{1,1},\dots,\hat a_{1,n})$
satisfying Property~\ref{indep_functions}.
\newline
Generators:
\be\label{gen_Scheme_A}
\begin{array}{l}
\hat a_{0,i} \hat a_{1,i}  \;\;\;  i\in \{1,\dots n\} \\
\hat a_{0,i} \hat a_{0,j}  \;\;\;  (i,j)\in \Gamma.
\end{array}
\ee
\begin{definition}
A diagonal model  is defined as Model A with $\Gamma=\emptyset$.
\end{definition}
Clearly, an diagonal model defines an algebraic set with $2^n$ isolated
points. Each point satisfies the linear equations
\be
\hat a_{s_i,i}=0, \;\;\;  i\in \{1,\dots,n\}
\ee
for some $(s_1,\dots,s_n)\in\{0,1\}^n$.

If there is an algorithm with polynomial complexity 
and associated with Model~A, then it is possible to prove
that there is 
another algorithm with polynomial complexity and
associated with a diagonal model. More generally,
this formula reduction takes to a subexponential factoring
algorithm, provided that the parent algorithm outperforms 
the quadratic sieve algorithm. If the parent algorithm outperforms
the general number field sieve, then the reduced algorithm outperforms
the quadratic sieve. Thus, if we are interested to find a competitive
algorithm from Model~A, we need to search only the space of reduced formulas. 
If there is no algorithm outperforming the quadratic sieve with $\Gamma=\emptyset$,
then there is no algorithm outperforming the general number field 
for $\Gamma\ne \emptyset$.


\begin{theorem}
If there is a factoring algorithm  with subexponential asymptotic complexity
$e^{a (\log p)^\gamma}$ and associated with Model~A, then there is another algorithm 
associated with the diagonal model  with computational complexity upper-bounded by
the function $e^{\bar a (\log p)^\frac{\gamma}{1-\gamma}}$ for some $\bar a>0$.
In particular, if the first algorithm has polynomial complexity,
also the latter has polynomial complexity.
\end{theorem}
{\it Proof}.
Let us assume that the asymptotic computational complexity of the 
parent algorithm is $e^{a (\log p)^\gamma}$. For every $N_P$,
there is a Model~A  with $N_P$ isolated
points and generating a rational function ${\cal R}$ with complexity
${\bf C}_0(\xi)$ scaling as $e^{a (\log N_P)^\alpha}$ and a number of
parameters $M$ scaling as $(\log N_P)^\beta$, where $\gamma=\alpha/(1-\beta)$
and $0\le\beta<1$ (See~\ref{sec_complex}).
We denote by $\cal Z$ the set of isolated points.
Since the complexity ${\bf C}_0$ is lower-bounded
by a linear function of the dimension $n$, we have
\be\label{ineq_n_N0}
\log n\le a (\log N_P)^\alpha+O(1).
\ee
Let $\hat a_{0,1},\dots,\hat a_{0,n},\hat a_{1,1},\dots,\hat a_{1,n}$ 
be the set of linear functions representing the matroid and satisfying
Property~\ref{indep_functions}. The ideal generators are given by 
Eq.~(\ref{gen_Scheme_A}).

Let $m$ be the maximum number of functions in $\{\hat a_{0,i},\dots,\hat a_{0,n}\}$ 
which are simultaneously different from zero for  $\vec x\in \cal Z$. Thus,
we have that 
\be\label{bound_N0}
N_P\le \sum_{j=0}^m\frac{n!}{(n-j)! j!}\le \left(1+n^{-1}\right) {n}^{m} .
\ee
There is a point $\vec x_2$ in $\cal Z$ such that 
$\hat a_{0,1},\dots,\hat a_{0,m}$ are different from zero and
$\hat a_{0,m+1}=\hat a_{0,m+1}=\dots=\hat a_{0,n}=0$,
up to permutations of the indices.

Let us set these last $n-m$ functions equal to zero by dimension reduction. 
The new set of generators is associated with another factoring algorithm 
(Lemma~\ref{lemma_dim_red}) and contains the clauses of the form
\be
\begin{array}{l}
\hat a_{0,i} \hat a_{1,i}  \;\;\;  i\in \{1,\dots,m\} \\
\hat a_{0,i} \hat a_{0,j}  \;\;\;  (i,j)\in \bar\Gamma\subseteq 
\{1,\dots,m\} \times \{1,\dots,m\}.
\end{array}
\ee
Since there is a point $\vec x_2$ such that $\hat a_{0,i}\ne$ for $i\in\{1,\dots,m\}$,
the set $\bar\Gamma$ turns out to be empty, so that the reduced model is diagonal.
The number of common zeros of the generators, say $N_1$, is equal to $2^m$. Using
Ineq.~(\ref{bound_N0}), we have that 
\be
(\log_2 N_1)(\log n)+\log\left(1+n^{-1}\right) \ge \log N_P.
\ee
Ineq.~(\ref{ineq_n_N0}) and this last inequality implies that
\be
\log N_P \le  K (\log N_1)^{\frac{1}{1-\alpha}}
\ee
for some constant $K$. Since the computational complexity, say $\bar {\bf C}_0$ of the rational 
function $\cal R$ associated with the reduced model is not greater than ${\bf C}_0$,
which scales as $e^{a (\log N_P)^\alpha}$,
we have that 
\be
\bar {\bf C}_0\le e^{\bar a (\log N_1)^\frac{\alpha}{1-\alpha}},
\ee
for some constant $\bar a$.
Similarly, since the number of parameters, say $\bar M$, of the reduced rational function
is not greater than $M$, we have that
\be
\bar M\le \bar K (\log \bar N_1)^\frac{\beta}{1-\alpha}
\ee
for some constant $\bar K$. Thus, the resulting factoring algorithm has a computational
complexity upper-bounded by 
$$
e^{\bar a(\log p)^\frac{\alpha}{1-\alpha-\beta} }=
e^{\bar a(\log p)^\frac{\gamma}{1-\gamma} }
$$
up to a constant factor.
The last statement of the theorem is proved in a similar fashion.
$\square$

The diagonal model with generators 
\be\label{simple_gene}
G_i=\hat a_{0,i} \hat a_{1,i},  \;\;\;\; \forall i\in\{1,\dots,n\} 
\ee
provides the simplest example of polynomials with an exponentially large number of
common zeros. The algebraic set ${\cal Z}={\cal Z}_P$ contains $2^n$ points, 
which are distinct because of Property~\ref{indep_functions}. This guarantees that 
the generated ideal is radical. Thus,  Hilbert's Nullstellensatz 
implies that every polynomial which is zero in ${\cal Z}$
can be written as $\sum_i F_i(\vec x) G_i(\vec x)$, where $F_1,\dots,F_n$ 
are polynomials (let us remind that $x_{n+1}=1$).

We impose that the polynomials $P_0,\dots,P_{n-M}$ are in the ideal 
generated by $G_1,\dots,G_n$, that is,
\be
P_k(\vec x)=\sum_i C_{k,i}(\vec x) \hat a_{0,i} \hat a_{1,i} \;\;\;  
\forall k\in\{0,\dots,n-M\}.
\ee
As there is no particular requirement on $P_0$, we can just set $C_{0,i}(\vec x)$ equal
to constants. In particular, we can take $P_0=\hat a_{0,1}\hat a_{1,1}$.
In this case, the unknown variables of Problem~\ref{problem1} are the
polynomials $C_{k,i}(\vec x)$ and 
the linear equations $\hat a_{s,k}$ under the constraints of Property~\ref{indep_functions}.
In the following section we tackle this problem with $C_{k,i}(\vec x)$ 
constant. 

\section{Quadratic polynomials}
\label{sec_quadr_poly}
In this section, we illustrate the procedure described previously by
considering the special case of $n-M+1$ quadratic polynomials 
in the ideal $I$ generated by the polynomials~(\ref{simple_gene}).
Namely, we take the polynomials $P_l$ of the form
\begin{equation}
\label{quad_poly}
P_l(\vec x)=\sum_{i=1}^n  c_{l,i}\hat a_{0,i}\hat a_{1,i}, \;\;\;\; 
l\in\{0,\dots,n-M\},
\end{equation}
where $c_{l,i}$ are rational numbers and the linear functions $\hat a_{s,i}$
satisfy Property~\ref{indep_functions}. Thus, there are  $2^n$ common rational 
zeros of the $n-M+1$ polynomials, which are also the zeros of the 
generators~(\ref{simple_gene}).
Each rational point is associated
with a vector $\vec s\in\{0,1\}^n$ so that the linear equations
$\vec a_{s_1,1}\cdot\vec x=0,\dots,\vec a_{s_n,n}\cdot\vec x=0$ are
satisfied. 

First, we consider the case with one parameter ($M=1$). We also assume that
all the $2^n$ rational points are in the parametrizable variety.
Starting 
from these assumptions, we end up to build a variety $\cal V$ with a number
$M$ of parameters equal to $n/2-1$ 
for $n$ even and $n\ge4$. Furthermore, we prove that there is no solution
with $M=1$ if $n>4$. We give a numerical example for $n=4$, which
takes to a rational function $\cal R$ with $16$ zeros. 
Then we build a parametrizable variety with a number of parameters equal to
$(n-1)/3$. Thus, the minimal number of parameters is some value between $2$ 
and $(n-1)/3$ for the considered model with the polynomials of the 
form~(\ref{quad_poly}).

\subsection{One parameter? ($M=1$)}
Given polynomials~(\ref{quad_poly}) and vectors $\vec a_{s,i}$ satisfying
Property~(\ref{indep_functions}), we search for a solution of Problem~\ref{problem1}
under the assumption $M=1$.
Let us first introduce some notations and definitions.
We define the $(n-1)\times n$ matrices
\begin{equation}
\label{def_matr_M}
{\bf M}^{\vec s}\equiv 
\left(
\begin{array}{ccc}
A_{1,1}^{(s_1)} & \dots & A_{1,n}^{(s_n)}    \\
\vdots & \ddots   & \vdots   \\
A_{n-1,1}^{(s_1)} & \dots & A_{n-1,n}^{(s_n)} 
\end{array}
\right),
\end{equation}
where
\begin{equation}
A_{k,i}^{(s)}\equiv\frac{\partial\hat a_{s,i}}{\partial x_k},
\end{equation}
The square submatrix of ${\bf M}^{\vec s}$ obtained by deleting the $j$-th column 
is denoted by ${\bf M}_j^{\vec s}$, that is,
\begin{equation}
{\bf M}_j^{\vec s}=
\left(
\begin{array}{cccccc}
A_{1,1}^{(s_1)} & \dots &  A_{1,j-1}^{(s_{j-1})} &  A_{1,j+1}^{(s_{j+1})} & \dots  &A_{1,n}^{(s_n)}    \\
\vdots & \ddots & \vdots & \vdots & \ddots  & \vdots   \\
A_{n-1,1}^{(s_1)} & \dots &  A_{n-1,j-1}^{(s_{j-1})} &  A_{n-1,j+1}^{(s_{j+1})} & \dots  &A_{n-1,n}^{(s_n)}   
\end{array}
\right),
\end{equation}
The vectors $\vec a_{0,i}$ and $\vec a_{1,i}$ are also briefly denoted by $\vec a_i$ and 
$\vec b_i$, respectively. Similarly, we also use the symbols $A_{k,i}$ and $B_{k,i}$
for the derivatives $A_{k,i}^{(0)}$ and $A_{k,i}^{(1)}$.

Problem~\ref{problem1} takes the specific form
\begin{problem}
\label{problem2}
Find coefficients $c_{l,i}$ and vectors $\vec a_{s,i}$ satisfying 
Property~(\ref{indep_functions}) such that
\bey
\label{prob2_1}
\sum_{i=1}^n c_{l,i} \left( A_{k,i}\vec a_i+B_{k,i}\vec b_i\right)=0
 \;\;\;\;  1\le k < l\le n-1, 
\vspace{2mm}
\\
\label{prob2_2}
\sum_{i=1}^n c_{l,i} A_{l,i} B_{l,i}=0  \;\;\;\; 1\le l\le n-1,
\vspace{2mm}
\\
\label{prob2_3}
\vec x\in{\cal Z}_P \Rightarrow 
\sum_{i=1}^n c_{l,i} \left( A_{l,i}\hat a_i+B_{l,i}\hat b_i\right)\ne 0
\;\;\;\; 1\le l\le n-1.
\eey
\end{problem}
Let us stress again that the problem is invariant with respect to 
the transformations~(\ref{poly_replacement},\ref{invar_trans}), the latter
taking to the transformation
\be
A_{k,i}^{(s)}\rightarrow A_{k,i}^{(s)}+\sum_{l=1}^{k-1}\bar\eta_{k,l}A_{l,i}^{(s)}
\ee
of the derivatives. 

We have the following.
\begin{lemma}
\label{lemma_inde_A}
For every $\vec s\in\{0,1\}^n$ and $j\in\{1,\dots,n\}$, the matrix ${\bf M}_j^{\vec s}$
has maximal rank, that is,
\begin{equation}
\det {\bf M}_j^{\vec s}\ne 0.
\end{equation}
\end{lemma}
{\it Proof}.
Let us prove the lemma by contradiction. There is a $j\in\{1,\dots,n\}$,
$l\in\{1,\dots,n-1\}$, and an $\vec s\in\{0,1\}^n$ 
such that the $l$-th row of ${\bf M}_j^{{\vec s}}$
is linearly dependent on the first $l-1$ rows. Thus,
there are coefficients $\lambda_1,\dots,\lambda_{l-1}$ such that
\begin{equation}
A_{l,i}^{(s_l)}+\sum_{k=1}^{l-1}\lambda_k A_{k,i}^{(s_k)}=0 \;\;\; \forall i\ne j
\end{equation}
With a change of variables of the form of Eq.~(\ref{invar_trans}), this equation
can be rewritten in the form
\begin{equation}
A_{l,i}^{(s_l)}=0 \;\;\; \forall i\ne j.
\end{equation}
Up to permutations $\hat a_i\leftrightarrow \hat b_i$, we have
\begin{equation}
\label{lin_dep_lemma}
B_{l,i}=0 \;\;\; \forall i\ne j.
\end{equation}
From Eq.~(\ref{prob2_2}), we have
$$
\sum_{i=1}^n c_{l,i} A_{l,i} B_{l,i}=0.
$$
From this equation and Eq.~(\ref{lin_dep_lemma}), we get the
equation $c_{l,j} A_{l,j} B_{l,j}=0$, implying that 
$c_{l,j} A_{l,j}=0$ or $c_{l,j} B_{l,j}=0$. Without loss of generality,
let us take
\be
\label{clne0}
c_{l,j} B_{l,j}=0.
\ee
Let $\vec x_0\in{\cal Z}_P$ be the vector orthogonal to $\vec b_1,\dots,\vec b_n$.
From Eq.~(\ref{prob2_3}) we have that
\be
c_{l,j} B_{l,j}(\vec a_j\cdot\vec x_0)\ne 0,
\ee
which is in contradiction with Ineq.~(\ref{clne0}).
$\square$

\begin{corollary}
\label{corol_c}
The coefficients $c_{n-1,i}$ are different from zero for every
$i\in\{1,\dots,n\}$.
\end{corollary}
{\it Proof}. 
Let us assume that the statement is false. Up to permutations, we have that
$c_{n-1,1}=0$. Lemma~\ref{lemma_inde_A} implies that there is an
integer $i_0\in\{2,\dots,n\}$ such that
$B_{n-1,i}=0$ for $i\notin\{1,i_0\}$, up to a transformation of the form of Eq.~(\ref{invar_trans}).
Thus,
\be
0=\sum_{i=1}^n c_{n-1,i}A_{n-1,i}B_{n-1,i}= c_{n-1,i_0} A_{n-1,i_0}B_{n-1,i_0},
\ee
the first equality coming from Eq.~(\ref{prob2_2}). Lemma~\ref{lemma_inde_A} also
implies that $A_{n-1,i_0}B_{n-1,i_0}\ne 0$ 
Thus, on one hand, we have that $c_{n-1,i_0}=0$.
On the other hand, we have
\be
c_{n-1,i_0} B_{n-1,i_0}(\vec a_{i_0}\cdot\vec x_0)\ne 0
\ee
from Eqs.~(\ref{prob2_3}),
where $\vec x_0$ is the vector orthogonal to $\vec b_1,\dots,\vec b_n$. 
Thus, we have a contradiction.  $\square$

Let us denote by ${\bf M}^{\vec s}_{j_1,\dots,j_m}$
the submatrix of ${\bf M}^{\vec s}$ obtained by deleting the last $m-1$ rows and
the columns $j_1,\dots,j_m$. 
Given the coefficient matrix
\be
{\bf c}\equiv
\left(
\begin{array}{ccc}
c_{0,1} & \dots & c_{0,n} \\
\vdots & \ddots & \vdots \\
c_{n-1,1} & \dots & c_{n-1,n}
\end{array}
\right),
\ee
let us define ${\bf c}_{j_1,\dots,j_m}$ as the $m\times m$ submatrix of
${\bf c}$ obtained by keeping the last $m$ rows and the columns 
$j_1,\dots,j_m$.

Lemma~\ref{lemma_inde_A} and Corollary~\ref{corol_c} are generalized by 
the following.
\begin{theorem}
\label{theorem_det}
For every $m\in\{1,\dots,n-1\}$, $\vec s\in\{0,1\}^n$, and $m$ distinct
integers $j_1,\dots,j_m\in\{1,\dots,n\}$,
the matrices ${\bf M}^{\vec s}_{j_1,\dots,j_m}$ and ${\bf c}_{j_1,\dots,j_m}$
have maximal rank, that is,
\bey
\label{det_reduc}
\det {\bf M}^{\vec s}_{j_1,\dots,j_m}\ne 0,  \\
\label{det_reduc2}
\det {\bf c}_{j_1,\dots,j_m}\ne 0.
\eey
\end{theorem}
{Proof.} The proof is by recursion. For $m=1$, the theorem comes from
Lemma~\ref{lemma_inde_A} and Corollary~\ref{corol_c}.
Thus, we just need to prove Eqs.~(\ref{det_reduc},\ref{det_reduc2}) by assuming that
\bey
\label{recur0}
\det {\bf M}^{\vec s}_{j_1,\dots,j_{m-1}}\ne 0, \\
\label{recur1}
\det {\bf c}_{j_1,\dots,j_{m-1}}\ne 0.
\eey
Let us first prove Eq.~(\ref{det_reduc}) by contradiction. If
the equation is false, then there is an $\vec s_0\in\{0,1\}^n$ and
$m$ distinct integers $i_1,\dots,i_m$ in $\{1,\dots,n\}$ such that
$\det {\bf M}^{\vec s_0}_{i_1,\dots,i_m}=0$.
By permutations, we can
set $i_h=h$. By suitable exchanges of $\hat a_i$ and $\hat b_i$, we can
set $s_i=1$ for every $i\in\{1,\dots,n\}$. 
There is an integer $l\in\{1,\dots,n-m\}$ such that $B_{l,i}=0$ for 
$i\in\{m+1,\dots,n\}$ up to a transformation of the form of Eq.~(\ref{invar_trans}).
From Eqs.~(\ref{prob2_1},\ref{prob2_2}), we have the $m$ equations
\be
\begin{array}{r}
\sum_{i=1}^m c_{l,i}A_{l,i}B_{l,i}=0  \\
\sum_{i=1}^m c_{n+1-m,i}A_{l,i}B_{l,i}=0  \\
\sum_{i=1}^m c_{n+2-m,i}A_{l,i}B_{l,i}=0  \\
\dots   \\
\sum_{i=1}^m c_{n-2,i}A_{l,i}B_{l,i}=0  \\
\sum_{i=1}^m c_{n-1,i}A_{l,i}B_{l,i}=0.
\end{array}
\ee
From Eq.~(\ref{recur0}), we have that $A_{l,i}\ne0$ and $B_{l,i}\ne0$ for some
$i\in\{1,\dots,m\}$, so that 
\be
\text{rank}
\left(
\begin{array}{ccc}
c_{l,1} & \dots & c_{l,m} \\
c_{n+1-m,1} & \dots & c_{n+1-s,m} \\
c_{n+2-m,1} & \dots & c_{n+2-s,m} \\
\vdots & \ddots & \vdots \\
c_{n-2,1} & \dots & c_{n-1,m}  \\
c_{n-1,1} & \dots & c_{n-1,m}
\end{array}
\right)< m.
\ee
Up to a transformation of the form of Eq.~(\ref{poly_replacement}),
there is an integer $l_0\in\{n+1-m,\dots,n-1\}\cup\{l\}$ such
that $c_{l_0,i}=0$ for $i\in\{1,\dots,m\}$. Eq.~(\ref{recur1}) implies
that $l_0=l$. Thus, $c_{l,1}=\dots c_{l,m}=0$, but this contradicts
Eq.~(\ref{prob2_3}) with $\vec x\in{\cal Z}_P$ orthogonal to $\vec b_1,\dots,\vec b_n$.

Let us now prove Eq.~(\ref{det_reduc2}) by contradiction. If the equation is false,
then there are $m$ distinct integers $i_1,\dots,i_m$ in $\{1,\dots,n\}$ such that
$\det {\bf c}_{i_1,\dots,i_m}=0$. Without loss of generality, let us take
$i_h=h$. Up to the transformation~(\ref{poly_replacement}),
there is an integer $l\in\{n-m,\dots,n-1\}$  such that $c_{l,i}=0$ for
$i\in\{1,\dots,m\}$. Eq.~(\ref{recur1}) implies that $l=n-m$. Thus,
\be
c_{n-m,1}=\dots=c_{n-m,m}=0.
\ee
Eq.~(\ref{det_reduc}) implies that there is an integer 
$i_0\in\{m+1,\dots,n\}$ such that $A_{n-m,i}=0$ for $i\in\{m+1,\dots,n\}\bs\{i_0\}$ 
up to transformation~(\ref{invar_trans}). Thus, we have from Eq.~(\ref{prob2_2})
that
\be
0=\sum_{i=1}^n c_{n-m,i}A_{n-m,i}B_{n-m,i}=
c_{n-m,i_0}A_{n-m,i_0}B_{n-m,i_0}.
\ee
Eq.~(\ref{det_reduc}) also implies that $A_{n-m,i_0}B_{n-m,i_0}\ne 0$, so that
$c_{n-m,i_0}=0$, which is in contradiction with Eq.~(\ref{prob2_3}) for $\vec x$
orthogonal to $\vec a_1,\dots,\vec a_n$. $\square$
\newline
In the following, this theorem will be used with $m\in\{1,2\}$.

Since all the coefficients $c_{n-1,i}$ are different from zero, we can set
them equal to $1$ by rescaling the vectors $\vec a_i$ or $\vec b_i$. 
Let us denote by $c_i$ the coefficients $c_{n-2,i}$. Theorem~\ref{theorem_det}
with $m=2$ implies that $c_i\ne c_j$ for $i\ne j$.
Eq.~(\ref{prob2_1}) with $l=n-1$ and $l=n-2$ takes the form
\bey
\label{nm2}
\frac{\partial }{\partial x_k}P_{n-1}=\sum_{i=1}^n \left( A_{k,i}\vec a_i+B_{k,i}\vec b_i\right)=0
 \;\;\;\;  1\le k \le n-2,  \\
\label{nm3}
\frac{\partial }{\partial x_k}P_{n-2}=
\sum_{i=1}^n c_i\left( A_{k,i}\vec a_i+B_{k,i}\vec b_i\right)=0
 \;\;\;\;  1\le k \le n-3.
\eey
These equations impose the form~(\ref{poly_affine}) to the last two
polynomials, $P_{n-1}$ and $P_{n-2}$, which must be independent from $n-2$ and
$n-3$ variables, respectively.
The first $n-2$ vector equations are linearly independent. Let us assume
the opposite. Then, there is a set of coefficients $\lambda_1,\dots,\lambda_{n-2}$
such that 
$\sum_{k=1}^{n-2}\lambda_k (A_{k,i},B_{k,i})=0$.
But this is impossible because of Property~\ref{indep_functions}. It also
contradicts Theorem~\ref{theorem_det}. The theorem also implies that
Eqs.~(\ref{nm3}) are linearly
independent. Since the vector space is $n+1$-dimensional, the vectors 
$\vec a_i$ and $\vec b_i$ must have $n-1$ vector constraints. Thus,
at least $n-4$ out of Eqs.~(\ref{nm3}) are linearly dependent on 
Eqs.~(\ref{nm2}).
First, let us show that $n-4$ is the maximal number of dependent equations.
Assuming the converse, we have
\be
c_i(A_{k,i},B_{k,i})=\sum_{l=1}^{n-2}\lambda_{k,l} (A_{l,i},B_{l,i}) \;\;\;\;\;
\forall k\in\{1,\dots,n-3\}.
\ee
for suitable coefficients $\lambda_{k,l}$. Let us define the linear 
superposition
\be
(A_i,B_i)\equiv \sum_{k=1}^{n-3} v_k (A_{k,i},B_{k,i})
\ee
with the coefficients $v_k$. Let $\bf \Lambda$ be the
$(n-2)\times(n-2)$ matrix with ${\bf \Lambda}_{k,n-2}=0$ 
and ${\bf \Lambda}_{k,l}=\lambda_{l,k}$
for $k\in\{1,\dots,n-2\}$ and  $l\in\{1,\dots,n-3\}$.
The coefficients 
$(v_1,\dots,v_{n-3})\equiv\vec v$ are defined by imposing
the $n-4$ constraints
\be
({\bf\Lambda}^s\vec v)_{n-2}=0 \;\;\;\;\; s\in\{1,\dots,n-4\}.
\ee
By construction, the pairs
\be\label{expo_form}
c_i^{k-1}(A_i,B_i)  \;\;\;\; k\in\{1,\dots,n-2\}
\ee
are linear superpositions of the derivatives $(A_{k,i},B_{k,i})$ with 
$k\in\{1,\dots,n-2\}$. Furthermore, the first $n-3$ pairs are 
linear superpositions of $(A_{k,i},B_{k,i})$ with 
$k\in\{1,\dots,n-3\}$. That is,
\be
\label{expo_deps}
\begin{array}{l}
c_i^{k-1} (A_i,B_i)=\sum_{l=1}^{n-3}\bar\lambda_{k,l}(A_{l,i},B_{l,i}) 
\;\;\;  k\in\{1,\dots,n-3\} \\
c_i^{n-3} (A_i,B_i)=\sum_{l=1}^{n-2}\bar\lambda_{n-2,l}(A_{l,i},B_{l,i})
\end{array}
\ee
for some coefficients $\bar\lambda_{k,l}$. From Lemma~\ref{lemma_inde_A}
and Corollary~\ref{corol_c} we have that the $n-2$ pairs~(\ref{expo_form})
are linearly independent. Indeed, Corollary~\ref{corol_c} implies that
$A_i\ne 0$ and $B_i\ne 0$ for every $i\in\{1,\dots,n\}$. Lemma~\ref{lemma_inde_A}
implies that $c_i^{k-1}$ are linearly independent for $k\in\{1,\dots,n-2\}$.
Equations~(\ref{expo_form},\ref{expo_deps}) can be also derived from Jordan's
theorem, Lemma~\ref{lemma_inde_A} and Corollary~\ref{corol_c}. See 
Appendix~\ref{lin_alg_tools}.

Thus, by a variable transformation, 
Eqs.~(\ref{nm2},\ref{nm3}) take the form
\be
\sum_{i=1}^n c_i^{k-1}\left( A_{i}\vec a_i+B_{i}\vec b_i\right)=0
\;\;\;\;\;\; k\in\{1,\dots,n-2\}.
\ee
and
\be
\frac{\partial (\hat a_i,\hat b_i)}{\partial x_k}=c_i^{k-1}(A_i, B_i) 
\;\;\;\;\;\; k\in\{1,\dots,n-2\}.
\ee
These equations imply that
$\sum_{i=1}^n d_i^{k-1} A_i B_i=0$ for $k\in\{1,\dots,2n-5\}$.
For $n>4$, we have in particular that
\be
\label{kill_AB}
\sum_{i=1}^n c_i^{k-1} A_i B_i=0 \;\;\;\; k\in\{1,\dots,n\}.
\ee
Since
\be
\det\left(
\begin{array}{ccc}
1   &   \dots   &    1  \\
c_1   &   \dots   &   c_n  \\
\vdots & \ddots & \vdots \\
c_1^{n-1} & \dots & c_n^{n-1}
\end{array}
\right)=\prod_{j>i} (c_j-c_i)
\ee
and $c_i\ne c_j$ for $i\ne j$, Eq.~(\ref{kill_AB}) implies that $A_i B_i=0$ for every 
$i\in\{1,\dots,n\}$. But this is in contradiction with Theorem~\ref{theorem_det}.

Thus, let us take exactly  $n-4$ out of Eqs.~(\ref{nm3}) linearly dependent on
Eqs~(\ref{nm2}). Let $\bar k$ be an integer in $\{1,\dots,n-3\}$ such that 
Eq.~(\ref{nm3}) with $k=\bar k$ is linearly independent of Eqs.~(\ref{nm2}). Thus, 
$$
c_i(A_{k,i},B_{k,i})=\bar\lambda_k c_i (A_{\bar k,i},B_{\bar k,i})+
\sum_{l=1}^{n-2}\lambda_{k,l}(A_{l,i},B_{l,i}) \;\;\;\; 
k\in\{1,\dots,n-3\}\bs\{\bar k\}.
$$
By a transformation of the first $n-3$ variables, we can rewrite this
equation in the form.
\be
c_i(A_{k,i},B_{k,i})=\sum_{l=1}^{n-2}\lambda_{k,l}(A_{l,i},B_{l,i}) \;\;\;\; 
k\in\{1,\dots,n-4\}.
\ee
By a suitable transformation of the first $n-2$ variables, 
the $n-2$ pairs $(A_{k,i},B_{k,i})$ can be split in two groups (see
Appendix~\ref{lin_alg_tools}), say, 
\be
\left.
\begin{array}{l}
\frac{\partial}{\partial x_k'} (\hat a_i,\hat b_i)  \equiv 
(A_{k,i}', B_{k,i}')=c_i^{k-1} (A_i',B_i') \;\;\;\; k\in\{1,\dots,n_1\} \\
\frac{\partial}{\partial x_k''}(\hat a_i,\hat b_i)  \equiv 
(A_{k,i}'', B_{k,i}'')=c_i^{k-1} (A_i'',B_i'') \;\;\;\; k\in\{1,\dots,n_2\}
\end{array} \right\}
\;\;\; n_1+n_2=n-2.
\ee
Equations~(\ref{nm2}) become
\begin{equation}
\label{nm2_vector_equation}
\begin{array}{l}
\sum_{i=1}^n c_i^{k-1}\left( A_i' \vec b_i+B_i' \vec a_i\right)=0 \;\;\; k\in\{1,\dots,n_1\} \\
\sum_{i=1}^n c_i^{k-1}\left(\bar A_i'' \vec b_i+\bar B_i'' \vec a_i\right)=0 \;\;\; k\in\{1,\dots,n_2\}.
\end{array}
\end{equation}
Given these $n-2$ vector constraints, all $n-2$ the derivatives 
$\partial P_{n-1}/\partial x_1',\dots,\partial P_{n-1}/\partial x_{n_1}'$,
$\partial P_{n-1}/\partial x_1'',\dots,\partial P_{n-1}/\partial x_{n_2}''$ are equal to
zero. Furthermore, we also have that 
$$
\begin{array}{l}
\frac{\partial}{\partial x_k'}P_{n-2}=0 \;\;\; k\in\{1,\dots,n_1-1\}  \\
\frac{\partial}{\partial x_k''}P_{n-2}=0 \;\;\; k\in\{1,\dots,n_2-1\},
\end{array}
$$
so that $P_{n-2}$ is independent of $n-4$ out of the $n-2$ variables 
$x_1',\dots,x_{n_1}$, $x_1'',\dots,x_{n_2}''$. Thus, we need to add
another vector equation such that 
$\left(w_1 \frac{\partial}{\partial x_{n_1}'}+w_2 \frac{\partial}{\partial x_{n_2}''}\right)P_{n-2}=0$
for some $(w_1,w_2)\ne (0,0)$. Up to a variable transformation, we can set $(w_1,w_2)=(1,0)$
so that the additional vector equation is 
\be
\label{additional_equation}
\sum_{i=1}^n c_i^{n_1}\left( A_i' \vec b_i+B_i' \vec a_i\right)
=0.
\ee
Equations~(\ref{prob2_2},\ref{nm2_vector_equation},\ref{additional_equation}) imply that
\bey
\label{eqs_coef_AB}
\sum_{i=1}^n c_i^{k-1} A_i' B_i'=0 \;\;\;  k\in\{1,\dots,2 n_1\},  \\
\label{eqs_coef_AbBb}
\sum_{i=1}^n c_i^{k-1} A_i'' B_i''=0 \;\;\;  k\in\{1,\dots,2 n_2\},  \\
\label{eqs_coef_AbB}
\sum_{i=1}^n c_i^{k-1} ( A_i' B_i''+A_i'' B_i')=0 \;\;\;  k\in\{1,\dots,n_1+n_2\}.
\eey
Since $A_i' B_i'$ and $A_i' B_i'$ are not identically equal to zero (as consequence of
Theorem~\ref{theorem_det}), the number of Eqs.~(\ref{eqs_coef_AB}) and Eqs.~(\ref{eqs_coef_AbBb}) 
is smaller than $n$, so that
$$
n_1\le \frac{n-1}{2},\;\;\; n_2\le \frac{n-1}{2}.
$$
Without loss of generality, we can assume that $n$ is even. Indeed, if Problem~\ref{problem1} 
can be solved for $n$ odd, then Lemma~\ref{lemma_dim_red} implies that it can be solve
for $n$ even, and {\it viceversa}. Since $n_1+n_2=n-2$, we have that
\be
n_1=n_2=\frac{n-2}{2}.
\ee
Let $W_1,\dots, W_n$ be $n$ numbers defined by the equations
\be
\label{def_W}
\sum_{i=1}^n c_i^{k-1} W_i=0   \;\;\;\; k\in\{1,\dots,n-1\}
\ee
up to a constant factor. Equations~(\ref{eqs_coef_AB},\ref{eqs_coef_AbBb},\ref{eqs_coef_AbB}) 
are equivalent to the equations
\bey
\label{AB}
A_i' B_i'=(k_0+k_1 c_i) W_i,   \\
\label{AbBb}
A_i'' B_i''=(r_0+r_1 c_i) W_i,   \\
\label{AbB}
A_i' B_i''+A_i'' B_i'=(s_0+s_1 c_i) W_i.
\eey
These equations can be solved over the rationals for the coefficients $c_i$, $B_i'$ and
$B_i''$ in terms of $A_i'$ and $A_i''$. The coefficients $c_i$ take a form which is
independent of $W_i$,
\be
c_i=\frac{r_0 A_i'^{\, 2}+k_0 A_i''^{\,2}-s_0 A_i' A_i''}{r_1 A_i'^{\, 2}+k_1 A_i''^{\,2}-s_1 A_i' A_i''},
\ee
so that we first evaluate $c_i$, then $W_i$ by Eq.~(\ref{def_W}) and, finally, $B_i'$ and
$B_i''$ by Eqs.~(\ref{AB},\ref{AbBb}).
It is possible to show that condition~(\ref{prob2_3}) for $l=n-1$ implies that $(k_1,r_1)\ne(0,0)$. 
Indeed, if $(k_1,r_1)=(0,0)$, then only half of the points in ${\cal Z}_P$ satisfies 
the inequality in the condition. 
Up to a variable change, we have 
$$
k_1\ne 0,\;\;\; s_1=0.
$$

Up to now we have been able to solve all the conditions of Problem~\ref{problem2} which
refer to the last two polynomials, that is, for $l=n-2,n-1$. The equations that need to
be satisfied are Eqs.~(\ref{nm2_vector_equation},\ref{additional_equation},
\ref{def_W},\ref{AB},\ref{AbBb},\ref{AbB}). Let us rewrite them all together.
\be
\boxed{
\begin{array}{c}
A_i' B_i'=(k_0+k_1 c_i) W_i, \;\;\;
A_i'' B_i''=(r_0+r_1 c_i) W_i   \\
A_i' B_i''+A_i'' B_i'=s_0 W_i,  \;\;\; k_1\ne 0
\vspace{1mm}   \\
\sum_{i=1}^n c_i^{k-1} W_i=0   \;\;\;\; k\in\{1,\dots,n-1\}
\vspace{1mm}   \\
\sum_{i=1}^n c_i^{k-1}\left( A_i' \vec b_i+B_i' \vec a_i\right)=0 \;\;\; k\in\{1,\dots,\frac{n}{2}\} \\
\sum_{i=1}^n c_i^{k-1}\left(\bar A_i'' \vec b_i+\bar B_i'' \vec a_i\right)=0 \;\;\; k\in\{1,\dots,
\frac{n}{2}-1\} 
\end{array}
}
\ee
Given $2n$ vectors $\vec a_1,\dots,\vec a_n,\vec b_1,\dots,\vec b_n$ satisfying these equations,
there are $n-1$ directions $\vec u_1,\dots,\vec u_{n-1}$ such that
\be
\begin{array}{l}
\vec u_{2k-1}\cdot \frac{\partial}{\partial\vec x} (\hat a_i,\hat b_i)=c_i^{k-1}(A_i',B_i') 
\;\;\; k\in\{1,\dots,\frac{n}{2}-1\}   \\
\vec u_{2k}\cdot \frac{\partial}{\partial\vec x} (\hat a_i,\hat b_i)=c_i^{k-1}(A_i'',B_i'') 
\;\;\; k\in\{1,\dots,\frac{n}{2}\}.
\end{array}
\ee
This can be easily verified by substitution.
Let us define the coordinate system  $(y_1,\dots,y_{n+1})\equiv \vec y$ such that
\be
\vec u_k\cdot \frac{\partial}{\partial\vec x} =\frac{\partial}{\partial y_k} \;\;\; 
k\in\{1,\dots,n-1\}.
\ee
Given the polynomials
\begin{equation}
\begin{array}{l}
P_{n-1}=\sum_{i=1}^n  \hat a_i\hat b_i \\
P_{n-2}=\sum_{i=1}^n  c_i \hat a_i\hat b_i 
\end{array}
\end{equation}
with $\hat a_i=\vec a_i\cdot \vec y$ and $\hat b_i=\vec b_i\cdot \vec y$, it is
easy to verify that
\begin{equation}
\begin{array}{l}
\frac{\partial P_{n-1}}{\partial y_k}=0, \;\;\;\; k\in\{1,\dots,n-2\}, \\
\frac{\partial P_{n-2}}{\partial y_k}=0, \;\;\;\; k\in\{1,\dots,n-3\}, \\
\frac{\partial^2 P_{n-2}}{\partial y_{n-2}^2}=0.
\end{array}
\end{equation}
The polynomial $P_{n-1}$ depends on $2$ variables (in the affine space) and
the polynomial $P_{n-2}$ depends linearly on an additional variable $y_{n-2}$.
Thus, the algebraic set of the two polynomials admits a Gaussian parametrization,
that is, the equations $P_{n-1}=0$ and $P_{n-2}=0$ can be solved with \emph{a la} Gauss 
elimination of two variables. Note that the polynomial $P_{n-1}$ has rational 
roots by construction.
The next step is to satisfy the conditions of Problem~\ref{problem2} for the other
polynomials 
$P_{1},\dots,P_{n-3}$  by setting  $c_{k,i}$ and the other remaining free coefficients.
It is interesting to note that it is sufficient to take $c_{2s,i}=c_i^{n/2-s}$ with 
$s\in\{1,\dots,(n-4)/2\}$ for satisfying every condition of Problem~\ref{problem2}
for $l$ even. The polynomials $P_2,P_4,\dots,P_{n-2}$ take the form
\be\label{poly_even}
P_{2s}=\sum_{i=1}^n c_i^{n/2-s} \hat a_i\hat b_i, \;\;\; s\in\{1,\dots,(n-4)/2\}.
\ee
Furthermore, we can choose $c_{1,i}$ such that $\partial^2 P_1/\partial x_1^2=0$.
With this choice, we have that 
\be
\left.
\begin{array}{l}
\frac{\partial P_l}{\partial y_k}=0, \;\;\;\; k\in\{1,\dots,l-1\} \\
\frac{\partial^2 P_l}{\partial y_l^2}=0
\end{array}
\right\} \;\;\; l\in\{2,4,\dots,n-4,n-2\}\cup\{1,n-1\}.
\ee
Thus, we are halfway to solve Problem~\ref{problem2}, about half of the conditions
are satisfied. The hard core of the problem is to solve the conditions for 
$P_1,P_3,\dots,P_{n-3}$. 
The form of Polynomials~(\ref{poly_even}) is not necessarily the most general. Thus, let
us take a step backward and handle Problem~\ref{problem2} for the polynomial $P_{n-3}$
with the equations derived so far. We will find that this polynomial cannot
satisfy the required conditions if $n>4$, so that the number of parameters has
to be greater than $1$.

Let us denote by $d_i$ the coefficients $c_{n-3,i}$. Eqs.~(\ref{prob2_1},\ref{prob2_2})
with $l=n-3$ give the equations
$$
\sum_{i=1}^n e_i \left(A_{k,i} B_{k',i}+A_{k',i} B_{k,i}\right)=0 \;\;\; 
k,k'\in\{1,\dots,n-3\},
$$
which imply that
$$
\begin{array}{l}
\sum_{i=1}^n e_i c_i^{k+k'-2} A_i' B_i'=0\;\;\; k,l\in\{1,\dots,\frac{n}{2}-1\} \\
\sum_{i=1}^n e_i c_i^{k+k'-2} A_i'' B_i''=0\;\;\; k,l\in\{1,\dots,\frac{n}{2}-2\} \\
\sum_{i=1}^n e_i c_i^{k+k'-2} \left( A_i' B_i''+ A_i'' B_i'\right)  =0\;\;\; 
\left\{
\begin{array}{l}
k\in\{1,\dots,\frac{n}{2}-1\} \\
k'\in\{1,\dots,\frac{n}{2}-2\}
\end{array}
\right.
\end{array}
$$
that is,
\be
\label{eqs_for_e}
\begin{array}{l}
\sum_{i=1}^n e_i c_i^{k-1} A_i' B_i'=0\;\;\; k\in\{1,\dots,n-3\} \\
\sum_{i=1}^n e_i c_i^{k-1} A_i'' B_i''=0\;\;\; k\in\{1,\dots,n-5\} \\
\sum_{i=1}^n e_i c_i^{k-1} \left( A_i' B_i''+ A_i'' B_i'\right)  =0\;\;\; 
k\in\{1,\dots,n-4\}.
\end{array}
\ee
These equations imply that
\be
\begin{array}{l}
e_i A_i' B_i'=F_{11}(c_i) W_i  \\
e_i A_i'' B_i''=F_{22}(c_i) W_i  \\
e_i \left(A_i' B_i''+ A_i'' B_i'\right) =F_{12}(c_i) W_i,
\end{array}.
\ee
where $F_{11}(x)$, $F_{22}(x)$ and $F_{12}(x)$ are polynomials of degree lower than $3$, $5$ and $4$,
respectively. Thus,
\be
e_i=\frac{F_{11}(c_i)}{k_0+k_1 c_i}=\frac{F_{22}(c_i)}{r_0+r_1 c_i}=
\frac{F_{12}(c_i)}{s_0}.
\ee
The second and third  equalities give polynomials of degree lower than $6$ and $5$, 
respectively. Since $c_i\ne c_j$ for $i\ne j$ and $n$ is even, the coefficients of
these polynomials are equal to zero for $n>4$. In particular, $k_0+k_1 c_i$ divides
$F_{11}(c_i)$ and, thus, $e_i$ is equal to a linear function of $c_i$. We have that
$P_{n-3}=q_1 P_{n-2}+q_2 P_{n-1}$ for some constants $q_1$ and $q_2$, so that
there is no independent polynomial $P_{n-3}$ satisfying the required conditions
for $n>4$. 
In conclusion, we searched for a solution of Problem~\ref{problem2} with one parameter ($M=1$), but
we ended up to find a solution with $n/2-1$ parameters. Let us stress that we have not
proved that $M$ cannot be less than $n/2-1$, we have only proved that $P_{n-3}$ 
cannot satisfy the required conditions, so that solutions with $M>1$ may exist.
Furthermore, we employed the condition $M=1$ in some
intermediate inferences. Thus, to check the existence of better solutions, we need to consider 
the case $M\ne 1$ from scratch.

For the sake of completeness, let us write down the solution for $n=4$. Eqs.~(\ref{eqs_for_e})
reduce to 
\be
\sum_{i=1}^n e_i A_i' B_i'=0,
\ee
Up to a replacement $P_{1}\rightarrow \lambda_1 P_1+\lambda_2 P_2+\lambda_3 P_3$ for some constants
$\lambda_i$ with $\lambda_1\ne 0$, we have that
\be
e_i=\frac{1}{k_0+k_1 c_i}.
\ee
Thus, the $4$ polynomials take the form
\be
\begin{array}{ll}
P_0=\hat a_1\hat b_1, \;\;\;\;
& P_1=\sum_{i=1}^{4}\frac{\hat a_i \hat b_i}{k_0+k_1 c_i} \\
P_2=\sum_{i=1}^{4}c_i \hat a_i \hat b_i  \;\;\;\;
& P_3=\sum_{i=1}^{4} \hat a_i \hat b_i.
\end{array}
\ee
Let us give a numerical example with $4$ polynomial, built by using the derived equations.
\subsubsection{Numerical example with $n=$4}
Let us set $A_i'=i$, $A_i''=1$, $k_0=k_1=r_0=1$, $r_1=2$, and $s_0=3$. Up to
a linear transformation of $x_3$ and $x_4$, this setting gives the polynomials
\be
\begin{array}{l}
P_3(x_3,x_4)=\\ 5 x_3 \left(8427 x_4+9430\right)-209 \left(3 x_4 \left(393 x_4+880\right)+1478\right) 
\vspace{1mm}  \\
P_2(x_2,x_3,x_4)= \\
5538425 x_3^2+18810 \left(1445 x_2+5718 x_4+6421\right) x_3-786258 \left(3 x_4 \left(267 x_4+598\right)+1004\right) 
\vspace{1mm} \\
P_1(x_1,x_2,x_3,x_4)= \\
2299 [205346285 x_3-38 (63526809 x_4+35594957)]- 5 [-2045057058 x_2^2+ \\
1630827 (1813 x_3+1254 x_4) x_2+2891872832 x_3^2+495958966272 x_4^2+   \\
4892481 x_1 (1254 x_2-1429 x_3-418)-87093628743 x_3 x_4]  \vspace{1mm}\\
P_0(x_1,x_2,x_3,x_4)= \\
\left(627 x_1+627 x_2-46 x_3+1881 \left(x_4+1\right)\right) \left(5016 x_1+6270 x_2+2555 x_3-3762 \left(4 x_4+5\right)\right)
\end{array}
\ee
Taking $x_4$ as the parameter $\tau$ and solving the equations $P_3=P_2=P_1=0$ with respect to $x_3$, $x_2$ and $x_1$,
we replace the result in $P_0$ and obtain, up to a constant factor,
\be
{\cal R}(\tau)=\frac{\prod_{k=1}^{16}(\tau-\tau_k)}{Q_1^2(\tau)Q_2^2Q_3^2(\tau)},
\ee
where 
\be
\begin{array}{l}
Q_1(\tau)=8427 \tau +9430,  \\
Q_2(\tau)=3 \tau  (393 \tau +880)+1478, \\
Q_3(\tau)=3 \tau  (9 \tau  (7 \tau 
   (5367293625 \tau +24273841402)+288165964484)+1954792734568)+1657527934720, \\
(\tau_1,\dots\tau_{16})= 
-\left(\frac{86}{69},\frac{800}{681},\frac{122}{105},\frac{3166}{2775},\frac{
   140}{123},\frac{718}{633},\frac{2452}{2163},\frac{5558}{4929},\frac{2578}{2
   289}, 
\frac{152}{135},\frac{1070}{951},\frac{3932}{3507},\frac{158}{141},
\frac{2072}{1851},\frac{1142}{1023},\frac{218}{201}\right)
\end{array}
\ee
Over a finite field $\mathbb{Z}_p$, one can check that the numerator has about $16$ distinct roots for $p\gg 16$.
For $p\simeq 16$, the roots are lower because of collision or because the denominator of some rational root $\tau_k$
is divided by $p$.
\subsubsection{Brief excursus on retro-causality and time loops}
Previously, we have built the polynomials~(\ref{poly_even}). Setting them equal to zero, we have
a triangular system of about $n/2$ polynomial equations that can be efficiently solved in $n/2$ variables,
say ${\bf x}_1$, given the value of the other variables, say ${\bf x}_2$. This system is more or less
symmetric, that is, the variables ${\bf x}_2$ can be efficiently computed given the first block 
${\bf x}_1$ (up to few variables). To determine the overall set of variables, we need the missing
$n/2$ polynomials in the ideal $I$.
It is possible to choose the coefficients $c_{l,i}$ of these polynomials in a
such a way that the associated equations have again a triangular form with respect to one of the
two blocks ${\bf x}_1$ and ${\bf x}_2$, up to few variables. Thus, we end up with two independent 
equations and a boundary condition,
\be
\begin{array}{l}
{\bf x}_2={\cal R}_1({\bf x}_1),  \\
{\bf x}_3={\cal R}_2({\bf x}_2),  \\
{\bf x}_3={\bf x}_1,
\end{array}
\ee
where ${\cal R}_1$ and ${\cal R}_2$ vectorial rational functions.
The first two equations can be interpreted as time-forward and time-backward processes. The last
equation identifies the initial state of the forward process with the final state of the backward
process. The overall process can be seen also as a deterministic process in a time loop.
This analogy is suggestive, since retro-causality is considered one possible explanation
of quantum weirdness. Can a suitable break of causality allow
for a description of quantum processes in a classical framework? To be physically interesting,
this break should not lead to a computational power beyond the power of quantum computers,
otherwise a fine tuning of the theory would be necessary to conceal, in a physical process,
much of the power allowed by the causality break. A similar fine tuning is necessary
if, for example,  quantum non-locality is explained with superluminar interactions. These classical 
non-local theories need an artificial fine tuning to account for non-signaling of quantum theory.

\subsection{$(n-1)/3$ parameters at most}
In the previous subsection, we have built a class of curves defined by systems of $n-1$ polynomial
equations such that about half of the variables can be efficiently solved over a finite field as
functions of the remaining variables. These curves and the polynomial $P_0$ have $2^n$ rational
intersection points. From a different perspective (discarding about $n/2$ polynomials), we have 
found a parametrizable variety with about
$n/2$ parameters such that its intersection with some hypersurface has $2^n$ rational points.

In this subsection, we show that the number of parameters can be dropped to about $n/3$ so
that about $2n/3$ variables can be efficiently eliminated, at least. In the following, we consider
space dimensions $n$ such that $n-1$ is a multiple of $3$. Let us define the integer
\be
n_1\equiv\frac{n-1}{3}.
\ee
Let us define the rational numbers $A_i,B_i,\bar A_i,\bar B_i$, $W_i$,  and $c_i$ with $i\in\{1,\dots,n\}$
as a solution of the equations
\be
\begin{array}{c}
A_i B_i=W_i,\;\;\; \bar A_i \bar B_i= W_i,  \\
A_i \bar B_i+\bar A_i B_i=2 c_i W_i,  \\
\sum_{i=1}^n c_i^{k-1} W_i=0 \;\;\;\; k\in\{1,\dots,n-1\},  \\
i\ne j \Rightarrow c_i\ne c_j.
\end{array}
\ee
The procedure for finding a solution has been given previously. 
We define the polynomials 
\be
P_s=\sum_i^n c_i^{s-1} \hat a_i \hat b_i, \;\;\;\; s\in\{1,\dots,n\}.
\ee
The linear functions $\hat a_i$ and $\hat b_i$ are defined by the $n-1$ linear equations
\be
\begin{array}{l}
\sum_{i=1}^{n_1} c_i^{k-1}(A_i \hat b_i+ B_i \hat a_i)=0, \;\;\; k\in\{1,\dots,n_1\} \\
\sum_{i=1}^{n_1} c_i^{k-1}(\bar A_i \hat b_i+\bar B_i \hat a_i)=0, \;\;\; k\in\{1,\dots,2 n_1\}.
\end{array}
\ee
These equations uniquely determine $\hat a_i$ and $\hat b_i$, up to a linear transformation
of the variables $x_i,\dots,x_{n+1}$.
Up to a linear transformation, we have
\be
\begin{array}{l}
\frac{\partial (\hat a_i,\hat b_i)}{\partial x_k}=c_i^{k-1} (\bar A_i,\bar B_i), \;\;\; k\in\{1,\dots,n_1\} \\
\frac{\partial (\hat a_i,\hat b_i)}{\partial x_{k+n_1}}=c_i^{k-1} (A_i,B_i), \;\;\; k\in\{1,\dots,n_1\}.
\end{array}
\ee
Since there are rational points in the curve, there is another variable, say $x_{2n_1+1}$, such that
the second derivative $\partial^2 P_1/\partial x_{2n_1+1}^2$ is equal to zero.
Using the above equations, we have
\be
\left.
\begin{array}{l}
\frac{\partial P_s}{\partial x_k}=0,  \;\;\;   k\in\{1,\dots,2 n_1-s+1\}, \\
\frac{\partial^2 P_s}{\partial x_k^2}=0,  \;\;\;   k=2 n_1-s+2.
\end{array}
\right\} \;\;\; s\in\{1,\dots,2 n_1+1 \}.
\ee
Thus, the first $2n_1+1=\frac{2n+1}{3}$ polynomials take the triangular form~(\ref{poly_affine}),
up to a reorder of the indices. These polynomials define a parametrizable variety with $(n-1)/3$
parameters. Stated in a different way, there is a curve and a hypersurface such that their
intersection contains $2^n$ points and at least $(2n+1)/3$ coordinates of the points in the curve 
can be evaluated efficiently given the value of the other coordinates. It is possible
to show that all the intersection points are in the parametrizable variety, that is,
they satisfy the third of Conditions~(\ref{gauss_constrs}).

\section{Conclusion and perspectives}
\label{conclusion}
In this paper, we have reduced prime factorization to the search of rational
points of a parametrizable variety $\cal V$ having an arbitrarily large number 
$N_P$ of rational points in the intersection with a hypersurface $\cal H$. 
To reach a subexponential factoring complexity, the number of parameters $M$ 
has to grow sublinearly in the space dimension $n$. In particular, 
If $N_P$ grows exponentially in $n$ and $M$ scales as a sublinear power
of $n$, then the factoring complexity is polynomial (subexponential) if
the computation of a rational point in $\cal V$, given the parameters, requires
a number of arithmetic operations growing polynomially (subexponentially) 
in the space dimension. Here, we have considered a particular kind of
rational parametrization. A set of $M$ coordinates, say $x_{n-M+1},\dots,x_n$, 
of the points in $\cal V$ are identified with the $M$ parameters,
so that the first $n-M$ coordinates are taken equal to rational functions of 
the last $M$ coordinates. In particular, the parametrization is expressed in a 
triangular form. The $k$-th variable is taken equal to a rational function 
${\cal R}_k={\cal N}_k/{\cal D}_k$ of the variables $x_{k+1},\dots,x_{n}$,
with $k\in\{1,\dots,n-M\}$. That is,
\be\label{triang_par}
\begin{array}{l}
x_k={\cal R}_k(x_{k+1},\dots,x_n),  \;\;\; k\in\{1,\dots,n-M\},
\end{array}
\ee
which parametrize a variety in the zero locus of the $n-M$ polynomials,
\be\label{triang_poly_form}
P_k={\cal D}_k x_k-{\cal N}_k, \;\;\;\; k\in\{1,\dots,n-M\}.
\ee
To reach polynomial complexity,
there are two requirements on these polynomials. First, they have to contain 
a number of monomials scaling polynomially in $n$, so that the computation of 
${\cal R}_k$ is efficient. For example, we could require that the degree is 
upper-bounded by some constant. Second, their zero locus has to share an
exponentially large number of rational points with some hypersurface $\cal H$
(a superpolynomial scaling $N_P\sim e^{b\,n^\beta}$ with $0<\beta<1$ is actually 
sufficient, provided
that the growth of $M$ is sufficiently slow).
The hypersurface is the zero locus of some polynomial $P_0$. Also the
computation of $P_0$ at a point has to be efficient.

We have proposed a procedure for building pairs $\{{\cal V},{\cal H}\}$
satisfying the two requirements. First, we define the set of $N_P$ rational
points. This set can depend on some coefficients. Since $N_P$ has to grow 
exponentially
in the dimension, we need to define them implicitly as common zeros of
a set of polynomials, say $G_1,G_2,\dots$. 
The simplest way is to take $G_k$ as
products of linear functions, like the polynomials~(\ref{quadr_polys}).
These polynomials generate an ideal $I$. The relevant 
information on the generators
is encoded in a satisfiability formula in conjunctive normal form without 
negations and a linear matroid. We have called these two objects a model.
Second, we search for $n-M$ polynomials
in $I$ with the triangular form~(\ref{triang_poly_form}). These polynomials
always exist. Thus, the task is to find a solution such that the polynomials
contain as few monomials as possible. 
This procedure is illustrated with the simplest example. The generators
are taken equal to reducible quadratic polynomials of the 
form~(\ref{quadr_polys}), whose associated algebraic set
contains $2^n$ rational points. We search for polynomials $P_k$
of the form $\sum_i c_i G_i$ with $c_i$ constant. First, we prove
that there is no solution for $M=1$ and space dimension greater than $4$.
Then, we find a solution for $M=(n-1)/3$. If there are solutions with
$M$ scaling sublinearly in $n$, then a factoring algorithm with polynomial
complexity automatically exists, since the computational complexity of 
the rational functions ${\cal R}_k$ is polynomial by construction.
The existence of such solutions is left as an open problem.

This work can proceed in different directions. First, it is necessary
to investigate whether the studied model admits solutions with
a much smaller set of parameters. The search has been performed in 
a subset of the ideal. Thus, if these solutions do not exist,
we can possibly expand this subset (if it is sufficiently large, there
is for sure a solution, but the polynomial complexity of ${\cal R}_k$ is
not guaranteed anymore). We could also
relax other hypotheses such as the distinguishibility 
of each of the $2^n$ rational points and their membership of
the parametrizable variety. More general ideals are another option.
In this context, we have shown that classes of models can be reduced
to smaller classes by preserving the computational complexity
of the associated factoring algorithms.
This reduction makes the search space smaller. It is interesting to
determine what is the minimal class of models obtained by this
reduction. This is another problem left open.
Apart from the search of better inputs of the procedure, there is a
generalization of the procedure itself. The variety $\cal V$ has
the parametrization~(\ref{triang_par}).
However, there are more general parametrizable
varieties which can be taken in consideration. 
It is also interesting to investigate if there is some
deeper relation with retro-causality, time loops and, possibly, a
connection with Shor algorithm. Indeed, in the
attempt of lowering the geometric genus of one of the non-parametrizable
curves derived
in the previous section, we found a set of solutions for the coefficients
over the cyclotomic number field, so that the resulting polynomials have
terms taking the form of a Fourier transform. Quantum Fourier transform
is a key tool in Shor's algorithm. This solution ends up to break
the curve into the union of an exponential large number of parametrizable
curves, thus it is not useful for our purpose. Nonetheless, the Fourier-like
forms in the polynomials remains suggestive.
Finally, the overall framework has some interesting relation with the
satisfiability problem.
Using a particular matroid, we have seen that there is a one-to-one correspondence
between the points of an algebraic set and the solutions of a satisfiability
formula (including also negations). To prove that a formula is satisfiable is
equivalent to prove that a certain algebraic set is not empty. This mapping
of SAT problems to an algebraic-geometry problem turns out to be a generalization
of previous works using the finite field $\mathbb{Z}_2$, see for example 
Ref.~\cite{hung}. It can be interesting to investigate whether part of the machinery 
introduced here can be used for solving efficiently some classes of SAT formulae.

\section{Acknowledgments}
This work was supported by Swiss National Science Foundation (SNF) and Hasler foundation
under the project n. 16057.

\appendix

\section{Basics of algebraic geometry}
\label{alg_geom}
\subsection{Ideals in polynomial rings and algebraic sets}
Given a field $K$, a \emph{polynomial ring} $K[X_1,\dots,X_n]$ is a ring whose
elements are polynomials in $n$ variables. An \emph{ideal} $I$ in the polynomial
ring is a subset of the ring closed with respect to the addition of elements in $I$ 
and the multiplication of elements in $I$ by elements of the ring. Every ideal~$I$ 
in the polynomial ring is generated by a finite set of polynomials in $I$ (Hilbert's
basis theorem). Given a set of generators $G_1,\dots,G_m$ of $I$, every element of 
the ideal can 
be written in the form $\sum_{k=1}^m F_k G_k$, where $F_k$ are polynomials in 
the ring. An ideal has infinite possible sets of generators. A
set of generators is a particular representation of the ideal and a change
of generators is similar to a change of basis in a vector space.
There is a correspondence between ideals and algebraic
sets. \emph{Algebraic sets} are subsets of an affine space whose elements are all 
the common zeros of some set of polynomials in the algebraically closed extension 
$\bar K$ of $K$. This set of polynomials generates an ideal.
By definition of ideals, the common zeros of a set of polynomials
are common zeros of every polynomial in the generated ideal. Thus, 
we can associate ideals with algebraic sets. This takes to the following
equivalent definition of algebraic set.
\begin{definition}
A subset $V$ of an affine space is an algebraic set if
there is an ideal $I$ such that the common zeros of the polynomials in
$I$ are all the elements in $S$. An algebraic set associated with an
ideal $I$ is denoted by ${\bf V}(I)$.
\end{definition}
This definition of algebraic set is more appealing as it does not refer to a 
particular representation of the ideal. The correspondence between ideals and 
algebraic sets is not one-to-one and an algebraic set is associated with many 
different ideals. For example, let ${\bf V}(I)$ be the algebraic set of the 
ideal $I$ with generators $G_1,\dots,G_m$. Let $J$ be the ideal
generated by $G_1^k, G_2,\dots,G_m$, where $k$ is some integer greater
than $1$. In general, the ideal $J$ is a strict subset of $I$. It is
easy to find examples for which $J\subset I$, such as $G_i=x_i$ with
$i\in\{1,\dots,n\}$ and $m=n$.
The common zeros of the polynomials in $J$ are all the elements in
${\bf V}(I)$. Thus ${\bf V}(I)={\bf V}(J)$, but $J\subset I$. 
A stricter correspondence between ideals and algebraic sets is obtained
by associating an algebraic set $V$ with the largest ideal $I$ such
that $V={\bf V}(I)$, called the \emph{vanishing ideal}.
\begin{definition}
Given an algebraic set $V$, we define the vanishing ideal of $V$ as
\be
{\bf I}(V)\equiv \bigcup_{{\bf V}(I)=V} I.
\ee
In other words, ${\bf I}(V)$ is the set of all the polynomials which are
zero in $V$.
\end{definition}
The composition ${\bf I}({\bf V}(I))\equiv I^*$ maps an ideal $I$ to
the largest ideal with same associated algebraic set. This map is idempotent,
that is, $(I^*)^*=I^*$. 
We have
$$
{\bf I}({\bf V}(I)) = {\bf I}({\bf V}(I^*))  =I^*.
$$
Thus, there is bijective correspondence between algebraic sets and
ideals of the form $I^*$.

Hilbert's Nullstellensatz identifies
the operation $I^*$ with the radicalization of $I$.
\begin{definition}
The \emph{radical} of an ideal $I$, denoted by $\sqrt{I}$, is an
ideal satisfying the double implication
$$
a\in \sqrt{I} \Leftrightarrow \exists k\mathbb\in {Z}^+ \text{  s.t.  } a^k\in I.
$$
The radical of an ideal $I$ contains $I$. If $\sqrt{I}=I$, $I$ is called
\emph{radical ideal} or \emph{semiprime ideal}.
\end{definition}
Also the radicalization of an ideal is an idempotent operation, that is, $\sqrt{\sqrt{I}}=\sqrt{I}$.
Since the polynomials $a$ and $a^k$ share the same zeros, it comes that
$$
{\bf V}(I)={\bf V}(\sqrt{I}).
$$
\begin{theorem} 
\label{hilbert_th}
(Hilbert's Nullstellensatz) Given the algebraic set $V\equiv{\bf V}(I)$ of an ideal $I$, a polynomial
is zero in $V$ if and only if it is an element of $\sqrt{I}$, that is, $\sqrt{I}=I^*$.
\end{theorem}
Thus,
$$
{\bf I}({\bf V}(I)) = {\bf I}({\bf V}(\sqrt{I}))  =\sqrt{I}.
$$
\begin{definition}
The \emph{intersection} of $I\cap J$ of two ideals $I$ and $J$ is an ideal
whose elements are both in $I$ and $J$.
\end{definition}
It comes from the definition of intersection that 
\be\label{ideal_inter}
{\bf V}(I\cap J)={\bf V}(I)\cup {\bf V}(J).
\ee
\begin{definition}
A \emph{prime ideal} $p$  of a polynomial ring $K[X_1,\dots,X_n]$ is a
strict subset of the ring such that, for every $a$ and $b$ in the ring, $a  b\in p$ 
implies that $a$ or $b$ are in $p$.
\end{definition}
A prime ideal generalizes the concept of prime integers. The ideal generated by
a prime integer in the ring $\mathbb{Z}$ is a prime ideal.
Excluding the ring among the prime ideals is like excluding $1$ among prime numbers, whose
ideal is the whole ring $\mathbb{Z}$.
\begin{lemma}
\label{rad_theorem}
The radical of an ideal $I$ is equal to the intersection of every
prime ideal containing $I$, that is,
\be
\sqrt{I}=\bigcap_{p \text{ prime},\; p\supseteq I} p
\ee
\end{lemma}
Let us illustrate this lemma with an example in one dimension.
Let $I$ be the ideal generated by
the polynomial $P=(x_1-k_1 x_0)^{a_1}\dots (x_1-k_m x_0)^{a_m}$, where the exponents
are positive integers and $i\ne j\Rightarrow k_i\ne k_j$. Its radical is generated
by $P_r= (x_1-k_1 x_0)\dots (x_1-k_m x_0)$. There are $m$  prime ideals 
containing $I$ and they are generated by the linear polynomials 
$(x_1-k_1 x_0), (x_1-k_2 x_0), \dots ,(x_1-k_m x_0)$. Their intersection is the ideal 
generated by $P_r$, in accordance with the lemma.

A direct consequence of Lemma~\ref{rad_theorem} is the following.
\begin{corollary}
If $I$ and $J$ are radicals, then also $I\cap J$ is radical.
\end{corollary}
This lemma and Eq.~(\ref{ideal_inter}) imply that the
order of the operations of intersection and radicalization can be exchanged, that is,
\be
\sqrt{I}\cap\sqrt{J}=\sqrt{I\cap J}.
\ee

\subsection{Product of ideals}
The product of two ideals $I$ and $J$ in the polynomial ring $K[X_1,\dots,X_n]$,
denoted by $I\cdot J$, is an ideal generated by the product of the generators of
$I$ and $J$. Namely, if $G_1,\dots,G_r$ and $H_1,\dots,H_s$ generates $I$ and $J$,
respectively, then the polynomials $G_i H_j$ with $i\in\{1,\dots,r\}$ and 
$j\in\{1,\dots,s\}$ generate $I\cdot J$. That is,
\be
I\cdot J\equiv\left\{\sum_{i,j} P_{i,j} G_i H_j|P_{i,j}\in K[X_1,\dots,X_n]\right\}.
\ee

It is easy to show that
\begin{equation}
{\bf V}(I \cdot J)={\bf V}(I)\cup{\bf V}(J)={\bf V}(I\cap J),
\end{equation}
that is, the algebraic set associated with $I\cdot J$ is equal to the union
of the algebraic sets associated with $I$ and $J$. Thus, the multiplication
of ideals acts on the associated algebraic sets like the intersection of the ideals.
The relation between product of radical ideals and their intersection is
given by the equation
\be
I, J \text{  radical ideals  }\Rightarrow I\cap J=\sqrt{I\cdot J}.
\ee
In general, we have
\begin{equation}
\sqrt{I}\cap\sqrt{J}=\sqrt{I\cap J}=\sqrt{I\cdot J}.
\end{equation}
On one side, multiplication of ideals has the advantage of being 
much simpler than intersection.
On the other side, the product of radicals is not generally a radical
ideal. As we will see, the intersection of ideals can have fewer generators 
with lower degree than the product. Furthermore, the intersection contains
all the polynomials that are zero in the associated algebraic set, which
can be a desired property.

\section{Linear-algebra tools}
\label{lin_alg_tools}
In this appendix, we discuss some direct consequences of Jordan's theorem that
are used in Sec.~\ref{sec_quadr_poly}.
Every $n\times n$ square matrix ${\bf M}$ can be transformed to a Jordan normal
form over an algebraically closed field through a basis change. This means that 
there is a basis of vectors grouped in $m$ sets $\{\vec v_{1,k}|k\in\{1,\dots,n_1\}\}$,
$\{\vec v_{2,k}|k\in\{1,\dots,n_2\}\},\dots$, $\{\vec v_{m,k}|k\in\{1,\dots,n_m\}\}$
with $\sum_{k=1}^m n_m=n$ such that the application of ${\bf M}$ acts as follows
\be
\left\{
\begin{array}{l}
\hat M\vec v_{s,k}=\lambda_s \vec v_{s,k}+\vec v_{s,k+1} \;\;\; k\in\{1,\dots,n_s-1\} \\
\hat M\vec v_{s,k}=\lambda_s \vec v_{s,k} \;\;\; k=n_s
\end{array}
\right.
\ee
In particular, if $n_s=1$ for every $s$, then the matrix is diagonalizable.
The existence of this basis is stated by Jordan's theorem. 

Let us apply Jordan's theorem to the following equation
\be\label{piv_eq}
{\cal A}\vec w_k= {\cal  B}\sum_{l=1}^{n} \lambda_{k,l}\vec w_l\;\;\;\; k\in\{1,\dots,n-\delta\},
\ee
where $\delta\in\{0,\dots,n-1\}$. The vectors $\vec w_1,\dots,\vec w_n$ are linearly independent,
on which the matrices $\cal A$ and $\cal B$ act.
The task is to find a set $\{\vec w_1,\dots,\vec w_n\}$ satisfying Eq.~(\ref{piv_eq}), given
$\cal A$ and $\cal B$.
In the following, we denote by boldface characters matrices acting
on the labels of the vectors $\vec w_k$. That is, $\bf A$ denotes an $i\times j$ matrix transforming
a set of vectors $\vec w_1,\dots,\vec w_j$ to the set of vectors
$\sum_l {\bf A}_{1,l}\vec w_l,\dots,\sum_l {\bf A}_{i,l}\vec w_l$.
The $n\times n$ matrices
acting on the vectors $\vec w_k$ are denoted by calligraphic characters.
The left-hand side of Eq.~(\ref{piv_eq}) has only vectors $\vec w_k$ with
$k\in\{1,\dots,n-\delta\}$. Thus, 
a transformation $\bf T$ on $\vec w_k$ preserves the form of the equation if the matrix 
has the form 
\be
{\bf T}=
\left(
\begin{array}{cc}
{\bf R} &  {\bf 0}  \\
{\bf S} &  {\bf D}  
\end{array}
\right),
\ee
where $\bf R$, $\bf S$, and $\bf D$ are $(n-\delta)\times(n-\delta)$, $\delta\times(n-\delta)$,
and $\delta\times\delta$ matrices, respectively. Let us define the basis of vectors
\be
\vec v_k\equiv \sum_{l=1}^n {\bf T}_{k,l} \vec w_l \;\;\;\;  k\in\{1,\dots,n\}.
\ee
Denoting by $\lambda$ the $(n-\delta)\times n$ matrix with elements ${\lambda}_{k,l}$,
the change of basis from $\vec w_k$ to $\vec v_k$ takes to the transformation
\be
{\bf\lambda}\rightarrow {\bf R}{\bf\lambda}{\bf T}^{-1}.
\ee
Jordan's theorem implies that there are $m$ sets of vectors 
$\{\vec v_{1,k}|k\in\{1,\dots,n_1\}\}$,
$\{\vec v_{2,k}|k\in\{1,\dots,n_2\}\},\dots$, $\{\vec v_{m,k}|k\in\{1,\dots,n_m\}\}$
such that
\be
\begin{array}{ll}
s\in\{1,\dots,\delta\}: &
{\cal A}\vec v_{s,k}= {\cal  B}\vec v_{s,k+1} \;\;\; k\in\{1,\dots,n_s-1\}    \\
s\in\{\delta+1,\dots,m\}:  &
\left\{
\begin{array}{l}
{\cal A}\vec v_{s,k}= {\cal  B}(\lambda_s\vec v_{s,k}+\vec v_{s,k+1}) \;\;\;  k\in\{1,\dots,n_s-1\}  \\
{\cal A}\vec v_{s,n_s}= {\cal  B}\lambda_s\vec v_{s,n_s}
\end{array}
\right.
\end{array}
\ee
with
\be
s\in\{1,\dots,\delta\} \left\{
\begin{array}{l}
\vec v_{s,k}\in\text{span}\{\vec w_1,\dots,\vec w_{n-\delta}\} \;\;\;
k\in\{1,\dots,n_s-1\}  \\
\vec v_{s,n_s}\notin\text{span}\{\vec w_1,\dots,\vec w_{n-\delta}\}.
\end{array}
\right.
\ee
If the matrix $\cal B$ is invertible, these equations provide a simple
way for building the vectors $\vec v_{s,k}$. We have that the vectors
take the form
\be
\left.
\begin{array}{ll}
s\in\{1,\dots,\delta\}: \;\;\; & \;\;
\vec v_{s,k}=({\cal B}^{-1}{\cal A})^{k-1} \vec v_s
\\
s\in\{\delta+1,\dots,m\}: \;\;\; &
\left\{
\begin{array}{l}
\vec v_{s,k}=({\cal B}^{-1}{\cal A}-\lambda_s)^{k-1} \vec v_s \\
({\cal B}^{-1}{\cal A}-\lambda_s)^{n_s} \vec v_s=0
\end{array}
\right.
\end{array}
\right\}
\;\;\; k\in\{1,\dots,n_s\},
\ee
where $\vec v_1,\dots,v_m$ are free vectors. Let us take
${\cal B}^{-1}{\cal A}$ is diagonalizable. Thus, the last line is equivalent
to the equation
$$
({\cal B}^{-1}{\cal A}-\lambda_s) \vec v_s=0\;\;\;  s\in\{\delta+1,\dots,m\}
$$
so that $\vec v_{s,k}=0$ for $k\in\{2,\dots,n_s\}$. Since the vectors 
$\vec v_{s,k}$ are linearly independent and, thus, different
from zero, we must have 
$$
n_s=1  \;\;\; s\in \{\delta+1,\dots,m\}.
$$
Thus, the general solution is given by the set of vectors 
$\{\vec v_{s,k}|s\in\{1,\dots,\delta\},k\in\{1,\dots,n_s\}\}$
and $\vec v_{\delta+1},\dots,\vec v_m$ defined by the equations
\be
\left\{
\begin{array}{ll}
s\in\{1,\dots,\delta\}:  & \;\;
\vec v_{s,k}=({\cal B}^{-1}{\cal A})^{k-1} \vec v_s 
\;\;\; k\in\{1,\dots,n_s\},  \\
s\in\{\delta+1,\dots,m\}:  & \;\;
({\cal B}^{-1}{\cal A}-\lambda_s) \vec v_s=0
\end{array}
\right.
\ee
with
$$
m-\delta+\sum_{s=1}^\delta n_s=n.
$$

\bibliography{biblio.bib}

\end{document}